\documentclass[a4paper,twocolumn,american,prl,superscriptaddress,longbibliography,fixfloat,notitlepage]{revtex4-2}
\usepackage{amssymb,graphicx,color}
\usepackage[intlimits]{amsmath}
\usepackage[english]{babel}
\usepackage[colorlinks]{hyperref}
\usepackage{footmisc}
\hypersetup{
colorlinks=blue,
linkcolor=blue,
filecolor=blue,
urlcolor=blue,
citecolor=blue
}

\usepackage{pdfpages} 
\usepackage{pgffor} 
\usepackage{xr} 

\makeatletter
\AtBeginDocument{\let\LS@rot\@undefined}
\makeatother

\def\supplementfilename{supp_mat4a}

\pdfximage{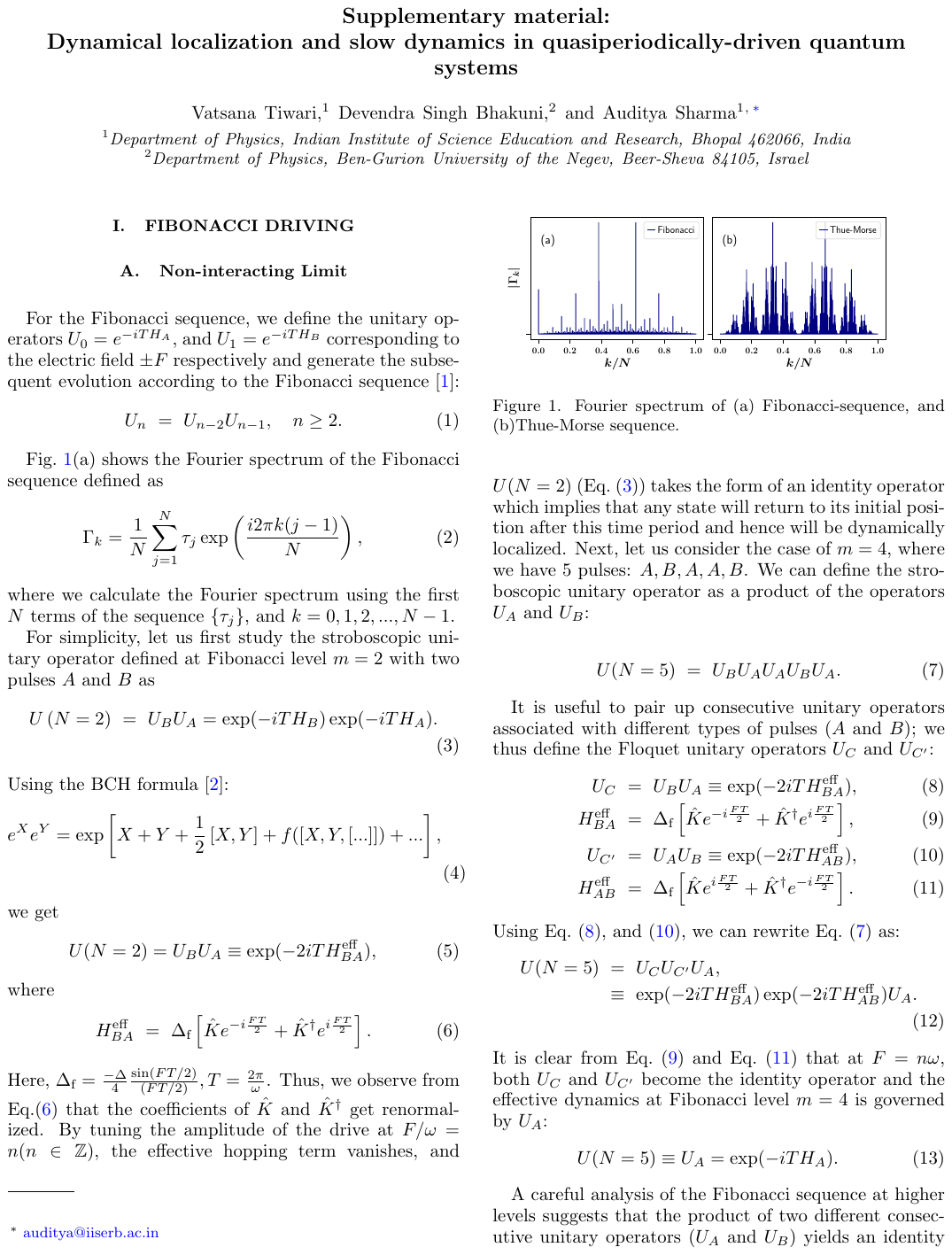}
\def\numbersupplementpages{\the\pdflastximagepages}

\newif\ifarXiv
\arXivtrue 

\pdfoutput=1
\usepackage[normalem]{ulem}
\usepackage{comment}
\usepackage{ragged2e}
\usepackage{enumitem}
\usepackage[table]{xcolor}
\usepackage{boldline}
\usepackage{cellspace}
\usepackage{array, multirow}
\usepackage{multirow, booktabs}
\usepackage{amsmath, amssymb,array}

\usepackage{graphicx}
\newcolumntype{P}[1
]{>{\centering\arraybackslash}p{#1}}
\definecolor{col0}{rgb}{0.75,0.75,0.75}
\definecolor{col1}{rgb}{0.9,0.92,0.9}
\definecolor{col2}{rgb}{0.92,0.94,0.85}
\definecolor{col3}{rgb}{0.75,0.83,0.93}
\begin{document}

\title{\hspace{-4pt}Dynamical localization and slow dynamics in quasiperiodically-driven quantum systems}
\author{Vatsana Tiwari}
\affiliation{Department of Physics, Indian Institute of Science Education and Research, Bhopal 462066, India}
\author{Devendra Singh Bhakuni}
\affiliation{Department of Physics, Ben-Gurion University of the Negev, Beer-Sheva 84105, Israel}
\author{Auditya Sharma}
\email{auditya@iiserb.ac.in}
\affiliation{Department of Physics, Indian Institute of Science Education and Research, Bhopal 462066, India}

\begin{abstract}	
  We investigate the role of a quasiperiodically driven electric field
  in a disordered fermionic chain. In the clean
  non-interacting case, we show the emergence of dynamical
  localization - a phenomenon previously known to exist only for a
  perfect periodic drive. In contrast, in the presence of disorder,
  where a high-frequency periodic drive preserves Anderson
  localization, we show that the quasiperiodic drive destroys it and
  leads to slow relaxation. Considering the role of interactions, we
  uncover the phenomenon of \emph{quasiperiodic driving-induced
  logarithmic relaxation}, where a suitably tuned drive (corresponding
  to dynamical localization in the clean, non-interacting limit) slows
  down the dynamics even when the disorder is small enough for the
  system to be in the ergodic phase. This is in sharp contrast to the
  fast relaxation seen in the undriven model, as well as the absence
  of thermalization (drive-induced many-body localization) exhibited
  by a high-frequency periodic drive.
\end{abstract}

\maketitle

\textit{Introduction}: The non-equilibrium properties of a quantum
system subjected to a time-dependent drive have been a topic of great
interest~\cite{grifoni1998driven,dalessio2014long,lazarides2014equilibrium,Lazarides2014periodic,bukov2015universal,eckardt2015high,ponte2015periodically,eisert2015quantum,khemani2016phase,ho2022quantum}. Some
notable phenomena associated with such driven systems include
dynamical localization in kicked
rotors~\cite{casati1979stochastic,grempel19841uantum,Lemarie2009observation},
dynamical
freezing~\cite{das2010exotic,roy2015fate,haldar2017dynamical,haldar2018onset, haldar2021dynamical,Haldar2022statistical},
Floquet topological
insulators~\cite{kitagawa2010topological,lindner2011floquet,nathananomalous2019},
Floquet
prethermalization~\cite{abanin2015exponentially,mori2016rigorous,abanin2017effective,abanin2017rigorous,weidinger2017floquet,beatrez2021floquet,peng2021floquet,fleckenstein2021prethermalization,fleckenstein2021prethermalization_1,morningstar2022simulation}
and time
crystals~\cite{else2016floquet,yao2017discrete,zhang2017observation,huang2018clean,khemani2019brief},
and Floquet many-body localization (MBL)
~\cite{lazarides2015fate,ponte2015many,abanin2016theory}. An
intriguing category of periodically driven systems involves an
\emph{electric-field drive}, which gives rise to a range of
fascinating phenomena, including dynamical
localization~\cite{dunlap1986dynamic,dunlap1988dynamic,lignier2007dynamical,eckardt2009exploring,bhakuni2018characteristic},
coherent destruction of Wannier Stark
localization~\cite{holthaus1995random,holthaus1995ac,bhakuni2018characteristic,tiwari2022noise,bhakuni2020drive},
and super Bloch
oscillations~\cite{kudo2011theoretical,longhi2012correlated}. Incorporating
many-body interactions further opens up fascinating possibilities,
such as Stark-MBL ~\cite{Schulz2019Stark,bhakuni2020entanglement,morong2021observation,guo2020stark,taylor2020experimental,evert2019bloch,bhakuni2020stability,ribeiro2020many,doggen2021stark,zisling2022transport},
drive-induced MBL~\cite{bairey2017driving,bhakuni2020drive}, and Stark time-crystals~\cite{kshetrimayum2020stark,liu2022discrete}. A crucial
question that arises is whether these characteristics persist in the
absence of a perfectly periodic drive.

While an enormous body of work has been devoted to periodic drives,
the exploration of the role of quasiperiodic driving has recently
gained traction~\cite{abal2002dynamical,verdeny2016quasi,
  nandy2017aperiodically,dumitrescu2018logarithmic,nandy2018steady,giergiel2019discrete,ray2019dynamics,maity2019fibonacci,else2020long,zhao2021random,mori2021rigorous,zhao2022localization,long2022many,martin2022effect,zhao2022suppression,zhao2022anomalous,das2023periodically}.
Features like
prethermalization~\cite{else2020long,zhao2021random,mori2021rigorous},
coherence restoration~\cite{mukherjee2020restoring}, and quasi-time
crystals~\cite{dumitrescu2018logarithmic,giergiel2019discrete,zhao2019floquet,zhao2021random} are associated with such
drives. Furthermore, experiments have realized dynamical phases
employing quasiperiodic driving~\cite{dumitrescu2022dynamical}. In this Letter,
we explore the properties of a system driven by a \emph{quasiperiodic
electric field}. Specifically, we address the question of whether a
quasiperiodic electric-field drive can give rise to dynamical
localization, and if it does, what the effect of interactions and
disorder on dynamical localization would be.



\begin{table}
\centering
\begin{tabular}{ |P{1.7cm}|P{2.0cm}|P{2.2cm}|P{2.4cm}| }
 \hlineB{2}
\rowcolor{col0}
\multirow{1.5}{*}{\textbf{Driving}} &\multirow{1.5}{*}{$\mathbf{W= 0}$} &\multirow{1.5}{*}{ $\mathbf{W\ne 0}$}& \multirow{1.5}{*}{ $\mathbf{W\ne 0,\ U\neq 0}$}\\
[2pt]     
\rowcolor{col0}

\multirow{1}{*}{\textbf{Protocol}}&\multirow{1}{*}{$\mathbf{U= 0}$}   &\multirow{1}{*}{$\mathbf{U= 0}$} & \multirow{1}{*}{\textit{ergodic regime}}\\
\hlineB{1.25}

  \rowcolor{col1}   
  
  & \multirow{1.7}{*}{Fast} &\multirow{1.7}{*}{Anderson} &\multirow{1.7}{*}{\hspace{-3ex} Fast} \\[5pt]\rowcolor{col1}  
 \multirow{-2.5}{*}{\textbf{Undriven}} & & &
 \\ \rowcolor{col1}
 & \multirow{-3.25}{*}{relaxation} &\multirow{-3.25}{*}{localization\cite{anderson1958absence}} &\multirow{-3.25}{*}{relaxation~\cite{agarwal2015anomalous}}\\ [-12pt]
   \hline\rowcolor{col2}
   
   & \multirow{1.5}{*}{Dynamical} & & \\
   [1pt]
\rowcolor{col2}
 \multirow{1.5}{*}{\textbf{Periodic}}  & &\multirow{1.5}{*} {Anderson} &\multirow{1.5}{*}{Drive-induced}\\
 [1pt]
\rowcolor{col2}
 &\multirow{-2.7}{*}{localization} & &\\
\rowcolor{col2}
  &\multirow{-2.75}{*}{\hspace*{-14pt}({\footnotesize Periodic}} &\multirow{-2.5}{*}{localization} &\multirow{-2.5}{*}{\hspace{2ex}MBL~\cite{bairey2017driving}} \\
\rowcolor{col2}
  &\multirow{-2.85}{*}{{\footnotesize oscillation})\cite{dunlap1986dynamic}} &\multirow{-2.75}{*}{~\cite{martinez2006delocalization} }&\\ [-9pt]
  \rowcolor{col2}
  \hline\rowcolor{col3}
  
   &\multirow{1.25}{*}{Dynamical}&\multirow{1.45}{*}{Slow}&\multirow{1.25}{*} {Drive-induced}\\[1pt]\rowcolor{col3}
     \multirow{1}{*}{\textbf{Quasi-}} &\multirow{1}{*}{localization}&\multirow{1}{*}{relaxation} &\multirow{1}{*}{logarithmic} \\[1pt]\rowcolor{col3}
\multirow{1}{*}{\textbf{periodic}}&\multirow{1.25}{*}{({\footnotesize Periodic}} &\multirow{1}{*}{$S(t)\propto t^\gamma,$} &\multirow{1}{*}{relaxation}  \\[-0.5pt]\rowcolor{col3}
  &\multirow{1}{*}{{\footnotesize oscillation})} &$\gamma<1$ &$S(t)\propto \log t$\\
    \hline
\end{tabular}
\caption{Schematic of the main results for the quasiperiodic drive and
  a comparison with existing results for the undriven and
  high-frequency periodic driving cases based on an analysis of the auto-correlation function and entanglement
  entropy. The third row contains the
  central findings of our work. Here, $ W, U$ refers to disorder and interaction strength, respectively.}
\label{table}
\end{table}
We address these questions by considering two discrete forms of
quasiperiodic driving: Fibonacci and
Thue-Morse~\cite{nandy2018steady,maity2019fibonacci,nandy2017aperiodically}. For
these sequences, we find that in the non-interacting limit, the
phenomenon of dynamical localization occurs when the parameters of the
drive are tuned appropriately. We derive the conditions under which it
is realized and numerically demonstrate this using the dynamics of
return probability and entanglement entropy. In the presence of
disorder, where the undriven system exhibits Anderson
localization~\cite{anderson1958absence,abrahams1979scaling}, we find
that the quasiperiodic drive destroys it and leads
to a slow relaxation of the return probability together with a
\textit{sub-linear} growth of entanglement entropy.

In the presence of many-body interactions, dynamical localization is
lost, and the system approaches the infinite-temperature
state. However, the approach is notably slower when the parameters are
specifically tuned at the dynamical localization point, as opposed to
arbitrary parameter choices. This follows from the fact that the drive
suppresses the hopping significantly at the dynamical localization
point. Next, we demonstrate how the hopping suppression can be
exploited to manipulate the dynamical behavior of a disordered
many-body interacting system. Starting from the ergodic phase in the
weak disorder limit of the undriven model, where the auto-correlation
function is known to exhibit power-law decay~\cite{agarwal2015anomalous}, we show that a quasiperiodic
electric-field drive can in fact lead to a slow logarithmic
relaxation. This is in sharp contrast to the drive-induced
MBL~\cite{bairey2017driving} that is seen for a
high-frequency periodic drive. In Table~(\ref{table}), the chief
findings from our work are summarized alongside already established
results in the literature.

\textit{Model Hamiltonian and driving protocol}: We consider a
disordered interacting one-dimensional tight-binding chain subjected
to a time-dependent electric field $\mathcal{F}\left(t\right)$. The model Hamiltonian can be written as:
\begin{eqnarray}
\label{eqn:eq1a}
H=-\frac{\Delta}{4}\sum_{j=0}^{L-2}\left(c_{j}^{\dagger}c_{j+1}+ h.c.\right)+\mathcal{F}\left(t\right)\sum_{j=0}^{L-1}j\left(n_{j}-\frac{1}{2}\right)\nonumber\\
 + \sum_{j=0}^{L-1}h_j\left(n_{j}-\frac{1}{2}\right) +U\sum_{j=0}^{L-2}\left(n_{j}-\frac{1}{2}\right)\left(n_{j+1}-\frac{1}{2}\right),\quad
\end{eqnarray}
where $c_{j}$ and $c_{j}^{\dagger}$ are fermionic annihilation and
creation operators respectively, $n_{j}$ are number operators, $\Delta
$ is the hopping strength, $U$ is the strength of nearest-neighbor
interaction, and $h_j$ is the on-site potential taken from a uniform
distribution: $h_j\in \left[-W,W\right]$. For the undriven case
($\mathcal{F}(t)=0$), Eqn.~\ref{eqn:eq1a} is the standard model of MBL
where a transition from an ergodic phase to an MBL phase occurs on
varying the disorder
strength~\cite{basko2006metal,vznidarivc2008many,luitz2015many,pal2010many},
although the precise value of the critical disorder strength is still
controversial~\cite{suntajs2020quantum,abanin2021distinguishing}. In
this work, we focus only on the ergodic side of the MBL
transition. For a periodic drive with time period $T$:
$\mathcal{F}(t+T)=\mathcal{F}(t)$, the clean non-interacting limit
($W=0, U=0$) exhibits dynamical localization for appropriately tuned
driving amplitude and
frequency~\cite{dunlap1986dynamic,dunlap1988dynamic,bhakuni2020drive},
while for a randomly fluctuating field there is no dynamical
localization~\cite{tiwari2022noise}. In this work, we focus on the
case where the time-dependent field is neither periodic nor random but
is rather quasiperiodic in nature and contains many frequencies in the
Fourier spectrum. We allow the field $\mathcal{F}$ to oscillate
between $\pm F$ after each period $T$ mimicing the Fibonacci and
Thue-Morse
sequences~\cite{nandy2017aperiodically,dumitrescu2018logarithmic,zhao2021random}.
\begin{figure}[b]
  \includegraphics[scale=0.365]{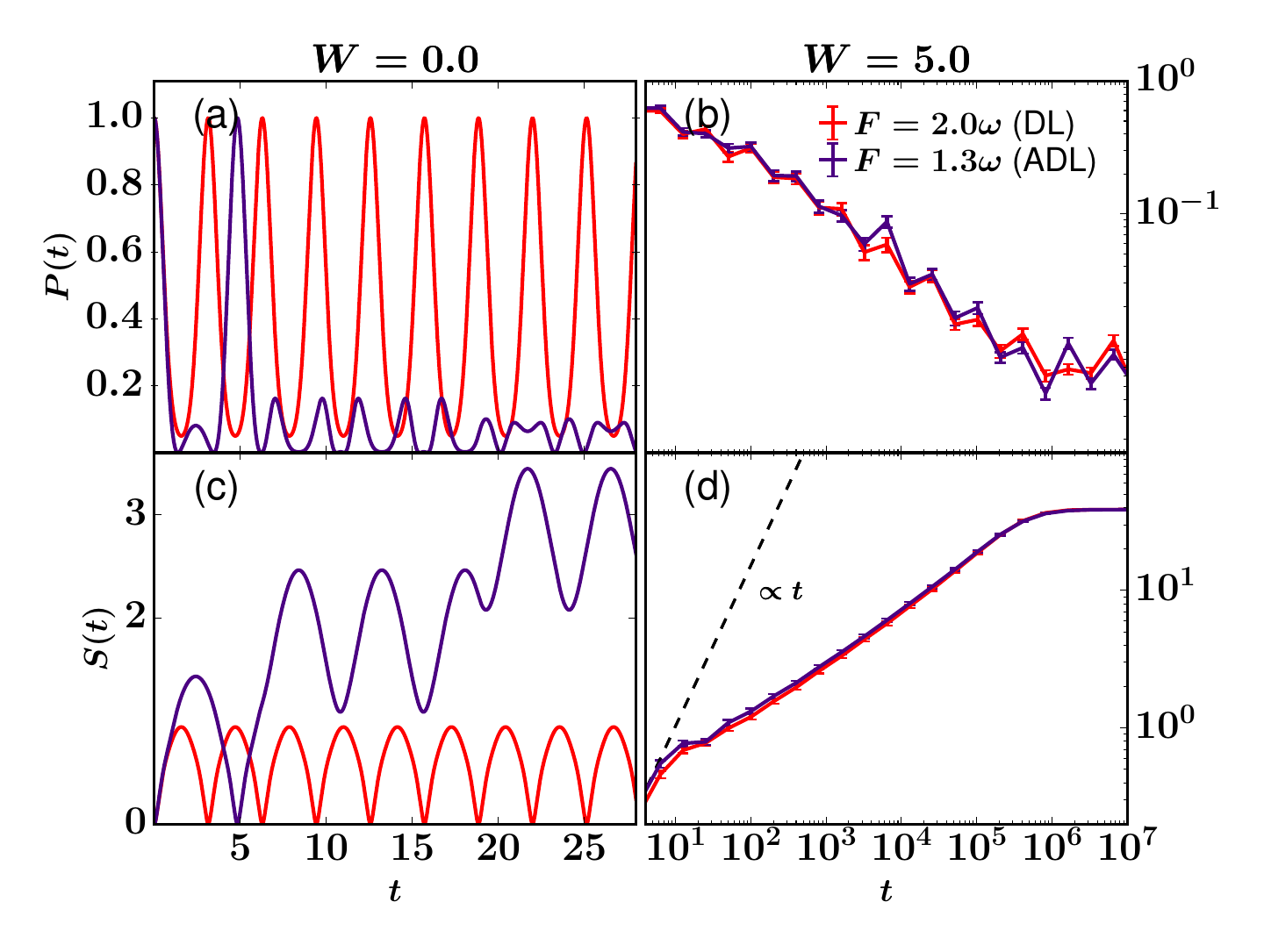}
  \caption{(a,b) Return probability $P(t)$, and (c,d) entanglement
    entropy $S(t)$ for a Thue-Morse driven clean ($W=0.0$) and
    disordered ($W=5.0$) system at dynamical localization (DL) ($F=
    2.0\omega $) and away from it (ADL) ($F=
    1.3\omega$). 
    The other parameters are $\Delta=4.0, \omega=1.0, L=200$. We take
    the time discretization $dt=0.001$ for (a,c). The data is averaged
    over $100$ realizations of disorder for (b,d). The black dashed
    line provide a guide to the ballistic growth.}
  \label{fig:fig3}
\end{figure}

In order to describe our driving protocol, it is convenient to
introduce the unitary operators $U_{B/A}=e^{-iTH_{B/A}}$ where
$H_{A/B}$ are the Hamiltonians corresponding to the field
$\mathcal{F}(t)=\pm F$ respectively (Eq.~\eqref{eqn:eq1a}).  For the
Fibonacci sequence, we start from the unitary operators $U_0 = U_A$
and $U_{1}=U_B$, and generate the subsequent evolution according to
the Fibonacci sequence~\cite{maity2019fibonacci}:
\begin{eqnarray}
U_{n}&=& U_{n-2}U_{n-1}, \quad  n\geq 2.
\end{eqnarray}
The Thue-Morse sequence (TMS), on the other hand, can be generated using the recurrence relation~\cite{zhao2021random}
\begin{eqnarray}
\label{tms}
U_{n+1}=\tilde{U}_{n}U_{n}, \quad \tilde{U}_{n+1}&=&U_{n}\tilde{U}_{n},
\end{eqnarray}
where we start with the unitary operators $U_{1}= U_{B}U_{A}$,
$\tilde{U}_{1}= U_{A}U_{B}$. The time evolution of an initial state
can be expressed as: $|\psi_n\rangle = U_n|\psi\left(0\right)\rangle$.

\begin{figure*}[t]
  \includegraphics[scale=0.426]{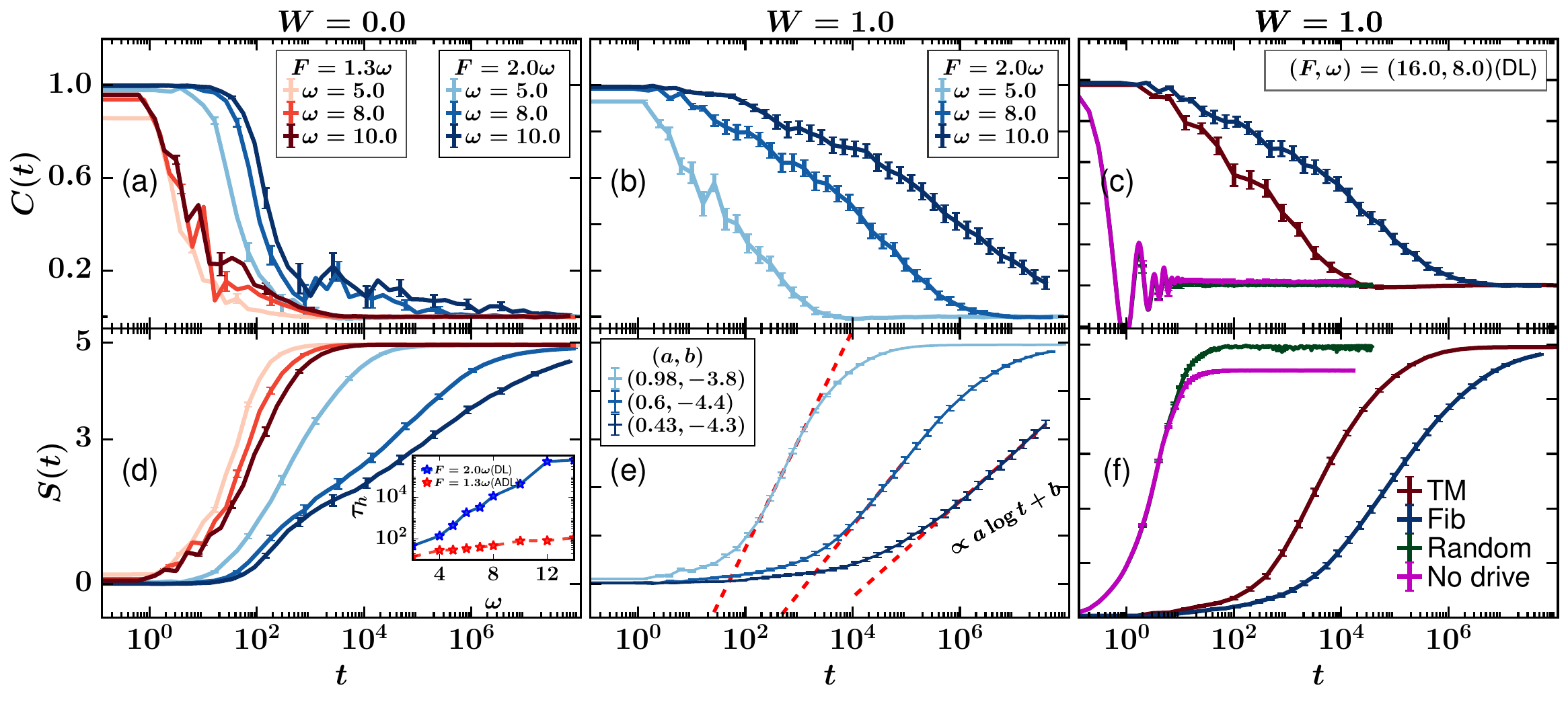}
  \caption{Dynamics of the auto-correlation function $C(t)$ and the
    half chain entanglement entropy $S(t)$ for the Fibonacci driven
    interacting fermionic system. (a, d): for $W=0.0$. The red shades
    correspond to a tuning of the parameters away from dynamical
    localization ($F=1.3\omega$, ADL), and blue shades correspond to a
    tuning at dynamical localization ($F=2.0\omega $, DL). The inset
    shows the heating time $\tau_{h}$ as a function of $\omega $. (b,
    e): for $W=1.0$. When disorder is present, we have averaged over
    $100$ disorder realizations. In (e), the red dashed
      lines are fits to the function $a\log t+b$. (c, f): Comparison
    of dynamics of differently driven and undriven interacting
    disordered systems. The other parameters are $\Delta=4.0, U=1.0,
    F=2.0\omega$, and $L=16$.}
  \label{fig:fig4}
\end{figure*}
\textit{Dynamical localization}: We first consider the clean
non-interacting limit ($W=0,\ U=0$). In this case, the Hamiltonian
\eqref{eqn:eq1a} can be written in terms of the unitary
operators:~\cite{hartmann2004dynamics}
$\hat{K}=\sum_{n=-\infty}^{n=\infty}|n\rangle \langle
n+1|,\ \hat{K}^{\dagger}=\sum_{n=-\infty}^{n=\infty}|n+1\rangle
\langle n\mid,\ \text{and}\ \hat{N}=\sum_{n=-\infty}^{\infty}n|
n\rangle \langle n|,$ as
\begin{eqnarray}
\hat{H}=-\frac{\Delta}{4}\left(\hat{K}+\hat{K}^{\dagger}\right)+\mathcal{F}\left(t\right)\hat{N}.
\end{eqnarray}
These operators follow the commutation relations: $[\hat{K},
  \hat{N}]=\hat{K}$, $[\hat{K}^{\dagger},
  \hat{N}]=-\hat{K}^{\dagger}$, $[\hat{K}, \hat{K}^{\dagger}]=0.$ With
this form of the Hamiltonian and the commutation relations, we can
write down the effective Hamiltonian $H_{\text{eff}}$. For Thue-Morse
driving, we work out an expression for the stroboscopic unitary
operator defined at the $m^{th}$ Thue-Morse level:
$U\left(N=2^{m}\right)$. With the aid of Eq.~\eqref{tms}, we can
express $U$ as a product of a string of the unitary operators $U_{A}$
and $U_{B}$, which using the Baker–Campbell–Hausdorff
formula~\cite{Hall2015},
can be written in terms of
$H_{\text{eff}}$ as~\cite{supplementary}
\begin{eqnarray}
\label{eqn:H}
U(N=2^{m})\equiv \exp(-i2^{m}TH_{\text{eff}}).
\end{eqnarray}
Here, $H_{\text{eff}}$ is given by
\begin{eqnarray}
\label{eqn:D}
H_{\text{eff}} \equiv  \Delta_{\text{eff}}\left(\hat{K}+\hat{K}^{\dagger}\right),\ \Delta_{\text{eff}}=-\frac{\Delta}{4}\left[\frac{\sin \left(2F\pi/\omega\right)}{\left(2F\pi/\omega\right)}\right],\quad 
\end{eqnarray}
where, $\omega=\frac{2\pi}{T}$. Thus we see from Eq.~\eqref{eqn:D}
that the coefficients of $\hat{K}$ and $\hat{K}^{\dagger}$ get
renormalized. When the amplitude of the drive is tuned at
$\frac{F}{\omega}=\frac{n}{2}$($n \in \mathbb{Z}$),
$\Delta_{\text{eff}}$ vanishes and any state will return to itself
after a time period $T$, and thus, the system remains dynamically
localized. Similar results are obtained for Fibonacci driving~\cite{supplementary}. 

The return probability, which is a measure of the probability of
finding a particle at an initially localized site $n$ after a time $t$
is defined as $P(t)=|\langle\psi(0)|\psi(t)\rangle|^{2}$; in our
study, we take $n=L/2$. The dynamics of the return probability for the
Thue-Morse driven system is plotted in Fig.~\ref{fig:fig3}(a). For the
ratio $F/\omega = 2$, we find that the return probability has
oscillatory behavior. We have checked using stroboscopic dynamics that
it periodically returns to its original value of unity even for very
long times. On the other hand, tuning the ratio at $F/\omega = 1.3$,
we see that the return probability is vanishingly small. A similar
observation can be made when looking at the growth of entanglement
entropy defined
as~\cite{nielsen_chuang_2010,peschel2003calculation,roy2018entanglement}:
$ S(t)=-\text{Tr}(\rho_{L/2}\ln\rho_{L/2})$, where
$\rho_{L/2}=\text{Tr}_{1\leq i\leq L/2}\left\lbrace|\psi(t)\rangle
\langle \psi(t)|\right\rbrace$ is the reduced density matrix of half
the chain obtained by tracing out the other half of the
chain. Starting with an initial N\'{e}el state $|\psi(0)\rangle =
\prod_{i=1}^{L/2}\hat{c}^{\dagger}_{2i}|0 \rangle$, we study the
dynamics of the entropy in Fig.~\ref{fig:fig3}(c). Again, when
$F/\omega = 2$, we find oscillatory behavior, while for $F/\omega =
1.3$, $S(t)$ starts to grow in time. These observations confirm
the analytical prediction that when the ratio $F/\omega$ is tuned
properly, the system exhibits dynamical localization; else, it
behaves like a nearest-neighbor chain with renormalized hopping.

\textit{Slow dynamics in the disordered non-interacting system}: We
now consider the disordered case, which leads to Anderson localization
for any non-zero value of the disorder strength in the undriven
model~\cite{anderson1958absence}. We study the stroboscopic evolution
of the return probability and the entanglement entropy for disorder
strength $W=5.0$. From Fig.~\ref{fig:fig3}(b), we see no sign of
either dynamical or Anderson localization; instead, we observe that
Thue-Morse driving leads to a slow decay of the return probability
accompanied by a sub-linear growth of the entropy
$S(t)\propto t^{\gamma}$ where $\gamma<1$~\emph{}\footnote{\textit{Since the frequency of the drive considered is small, the data corresponding to DL and ADL seem to overlap; however, we have checked that for higher frequencies,
while a slow relaxation is generic, tuning at DL results in a further	slowing down of the dynamics }\textit{.}}. This is in stark contrast with the effect of a
high-frequency periodic drive, where Anderson localization remains
stable~\cite{hone1993locally,martinez2006delocalization}. 

  An understanding of the slow dynamics of a quasi-periodically driven
  system can be obtained by performing a high-frequency expansion for
  the first two cycles, for which the effective Hamiltonian can be
  written as
\begin{equation}\label{eq10:Heff_dis}
H_{\text{eff}}=\Delta_{\text{eff}}\lbrace \hat{K}e^{-iFT/4}+\hat{K}^{\dagger}e^{iFT/4}\rbrace+ D_{0}+H_\text{LRH},
\end{equation}
where $D_{0}$ is the static disorder term and $H_\text{LRH}$ contains
the longer-range hopping terms. A derivation of
Eqn.~\ref{eq10:Heff_dis} and the form of $H_\text{LRH}$ is provided in
Ref.~\cite{supplementary}. When dealing with higher order terms and a
higher level of the quasiperiodic sequence, the calculations become
more complex. However, valuable information can still be extracted
from $H_\text{eff}$ at this level. For frequencies larger than the
local bandwidth $\Delta_\xi$ of a system of size of the order of the
localization length $\xi$, $H_{\text{LRH}}$ can be neglected since it
contains factors involving powers of the time period $T$ which becomes
very small.  Moreover due to the renormalization $\Delta_\text{eff}$,
the effective disorder increases, i.e. $H_\text{eff}(J,W)\approx
H(J,W/J_\text{eff})$ and hence a stronger localization would be
expected. Thus $H_\text{eff}$ obtained for two cycles suggests the
stability of Anderson localization in the presence of a high frequency
drive. For a periodic drive, these two cycles repeat indefinitely;
therefore, indeed this robustness of Anderson localization in the
presence of high frequency periodic drive has been
reported~\cite{hone1993locally,martinez2006delocalization}.  For our
drive protocol, however, we must incorporate higher levels of the
quasiperiodic sequence. This introduces a mixture of low and high
frequency components. While frequencies larger than the local
bandwidth $\omega>>\Delta_\xi$ do not influence localization, lower
frequencies $\omega<<\Delta_\xi$ initiate transitions between the
localized states. This competition gives rise to a gradual relaxation
of $P(t)$ and a sub-linear growth of $S(t)$, ultimately leading to
delocalization.

\textit{Slow dynamics in the disordered interacting system}:
Having discussed the emergence of dynamical localization under
quasiperiodic driving, we now explore the interplay of quasiperiodic
driving, many-body interactions, and disorder. We employ exact
diagonalization for a system of size $L=16$ at half-filling and focus
on the dynamics of entanglement entropy and auto-correlation
function. For driven quantum systems, the entanglement entropy
typically saturates to the Page value: $S_\text{Page}=\frac{1}{2}(L\ln
2-1)$ which corresponds to the infinite temperature
state~\cite{page1993average}. The auto-correlation function is defined
as~\cite{agarwal2015anomalous,barlev2015absence,luitz2017ergodic,lezama2019apparent}
\begin{equation}
  C(t) = 4\langle \hat{S}^{z}_{L/2}(t) \hat{S}^{z}_{L/2}(0)\rangle,
  \label{eq:eq8}
\end{equation}   
where $\hat{S}^{z}_{i} = \hat{n}_{i} - \frac{1}{2}$. While in the
localized phase, the auto-correlation function $C(t)$ saturates to a non-zero
value, in the ergodic phase, it rapidly goes to
zero~\cite{agarwal2015anomalous,dumitrescu2018logarithmic}.

We first start with the clean limit $W=0$, and observe that dynamical
localization is destroyed in the presence of many-body
interactions. However, for Fibonacci driving (and also
Thue-Morse~\cite{supplementary}), the dynamics is found to be very
slow if the interactions are turned on at the dynamical localization
point as opposed to any arbitrary choice of the parameters.  In
Fig.~\ref{fig:fig4}(a,d), we plot the dynamics of auto-correlation
$C(t)$, and entanglement entropy $ S(t)$ for $\frac{F}{\omega}=2$ (at
DL) and $\frac{F}{\omega}=1.3$ (away from DL) and a range of driving
frequencies. We average the quantities over $50$ different initial
product states close to the N\'{e}el state. The entanglement entropy
rapidly reaches the Page value while the auto-correlation function
quickly decays to zero for $\frac{F}{\omega}=1.3$ for all the driving
frequencies considered. On the other hand, for $\frac{F}{\omega}=2$,
$C(t)$ exhibits a slow decay accompanied by a slow relaxation of
$S(t)$ to the Page value (Fig.~\ref{fig:fig4}(a,d)). On increasing the
driving frequency, the growth further slows down, suggesting slow
heating.

We next consider the disordered case with $W=1.0$, which in the
undriven model, lies in the ergodic region where $C(t)$ shows fast
decay~\cite{agarwal2015anomalous}. We plot the auto-correlation and
the entanglement entropy for Fibonacci driving in
Fig.~\ref{fig:fig4}(b,e) for different driving frequencies and with
the parameters tuned at the dynamical localization point. Although the
undriven system lies in the ergodic regime, we see that the
quasiperiodic drive induces logarithmically slow relaxation of the
auto-correlation and a slow growth of the entropy. Earlier work has
reported a logarithmic relaxation in the MBL phase
alone~\cite{dumitrescu2018logarithmic}.

\begin{figure}[t]
	\includegraphics[scale=0.22]{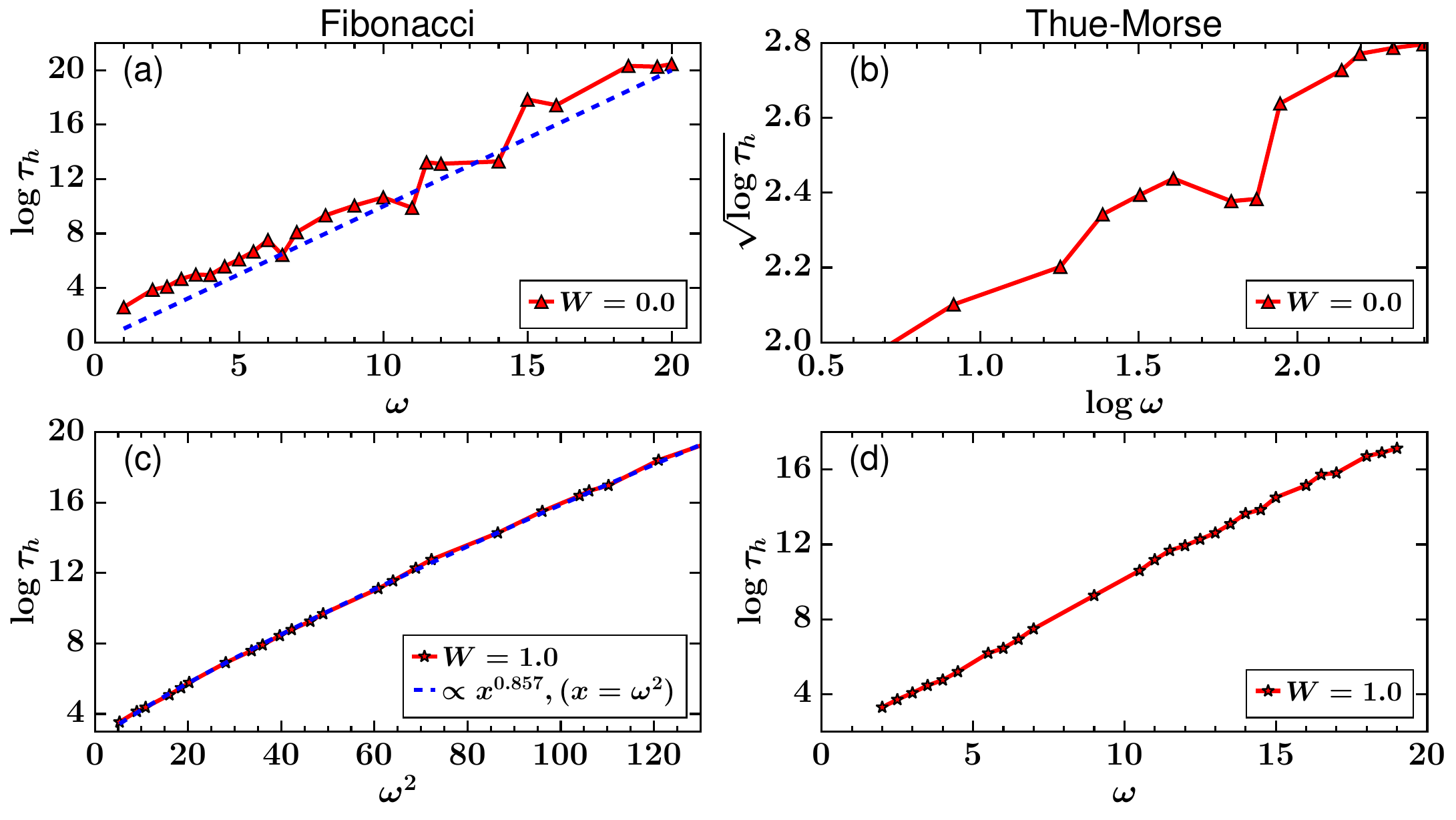}
	\caption{Heating time as a function of frequency for Fibonacci and Thue-Morse driving. (a,b) $W=0$. (c,d) $W=1$. The other model parameters are $J=1, U=1, F=2\omega$.}
	\label{fig:heating_time}
\end{figure}


 These findings can be understood through a
  high-frequency expansion analysis. Analogous to the non-interacting
  case, by considering the first two cycles, we get
  $H_\text{eff}(\Delta,W,U)\approx
  H(\Delta_\text{eff},W,U)$~\cite{supplementary}. This implies a
  suppression of hopping or, conversely, an augmented effective
  disorder strength $W/\Delta_{\text{eff}}$, driving the system
  towards the MBL regime~\cite{bairey2017driving,
    bhakuni2020drive}. However, when we incorporate higher terms of
  the quasi-periodic sequence, low-frequency components come into
  play, prohibiting the possibility of MBL. Nevertheless, a
  competition between the low and high-frequency components emerges,
  leading to slow relaxation of $C(t)$ and logarithmic growth of
  $S(t)$. The interplay between these components is also reflected in
  the lifetime of the slowly relaxing phase observed in the different
  driving protocols, as shown in Fig.~\ref{fig:fig4}(c, f). While a
  random drive with numerous low-frequency components yields fast
  relaxation, quasiperiodic driving results in a prolonged relaxation
  phase, with Fibonacci driving exhibiting even slower relaxation
  compared to Thue-Morse driving. This distinction can be attributed
  to the additional low-frequency components present in the Fourier
  spectrum of the Thue-Morse drive. A similar argument holds for the
  clean limit, where due to the hopping renormalization, we see a slow
  heating at dynamical localization as compared to away from it.

To investigate the impact of driving frequency on slow
  dynamics, we examined the heating time $\tau_h$ defined as the time
  when $S(t)$ attains half of the Page
  value~\cite{bhakuni2021suppression}, at different driving
  frequencies tuned at the dynamical localization point. We see that
  at the dynamical localization point, the heating time is several
  orders of magnitude greater than when the parameters are tuned away
  from it (Fig.~\ref{fig:fig4}(d) inset).  Fig.~\ref{fig:heating_time}
  illustrates the relationship between $\tau_h$ and $\omega$ for both
  Fibonacci and Thue-Morse driving, considering cases with $W=0.0$ and
  $W=1.0$. For $W=0.0$ (Fig.~ \ref{fig:heating_time}(a, b)), we
  observe an exponential dependence of $\tau_{h}$ on $\omega$ for
  Fibonacci driving, whereas Thue-Morse driving follows a
  sub-exponential ($\tau_{h}\sim e^{[\text{ln}\omega]^2}$) trend
  consistent with the theoretical bound~\cite{mori2016rigorous}. On
  the other hand, for $W=1.0$, where drive induces logarithmic
  relaxation, an extended heating time is evident. Fibonacci driving
  now displays a super-exponential dependence ($\tau_{h}\sim
  e^{\omega^\beta}, \beta>1$), while Thue-Morse driving exhibits an
  exponential dependence.

\textit{Conclusions}: We investigate the dynamics of fermions in a
disordered potential under the influence of a time-dependent electric
field generated from Fibonacci and Thue-Morse sequences. In the
absence of interactions and disorder, we demonstrate that dynamical
localization can be achieved through quasiperiodic driving and
identify the conditions for its realization. When disorder is
introduced, the quasiperiodic drive disrupts Anderson localization but
leads to a slow relaxation of observables. By introducing interactions
to a system near the dynamical localization point, we show a
significant suppression of heating compared to cases where parameters
deviate from it. Furthermore, utilizing the concept of hopping
suppression, we show a logarithmic relaxation induced by quasiperiodic
driving in the ergodic regime of an interacting system. This
drive-induced slow relaxation alters the dependence of heating time on
the driving frequency, resulting in a super-exponential dependence for
Fibonacci driving and an exponential dependence for Thue-Morse
driving, contrasting the expected exponential and sub-exponential
dependencies~\cite{dumitrescu2018logarithmic,mori2021rigorous}.

We look forward to our work inspiring ways to host non-equilibrium
phases of matter such as time
quasi-crystals~\cite{dumitrescu2018logarithmic,zhao2021random}. In
general, it would be intriguind to explore the possibility of using
quasiperiodic electric field drive to manipulate the properties of
quantum systems in the same spirit as a periodic drive. In the future,
it will be interesting to see features like coherent/incoherent
destruction of Wannier-Stark
localization~\cite{holthaus1995ac,bhakuni2019effect}, super-Bloch
oscillations~\cite{kudo2011theoretical,longhi2012correlated}, and the
effect of long-range interaction in such systems.

We are grateful to the High Performance Computing(HPC) facility at
IISER Bhopal, where large-scale calculations in this project were
run. We acknowledge QuSpin
package~\cite{weinberg2017quspin,weinberg2019quspin} for help in the
numerical exact-diagonalization. V.T is grateful to DST-INSPIRE for
her PhD fellowship. A.S acknowledges financial support from SERB via
the grant (File Number: CRG/2019/003447), and from DST via the
DST-INSPIRE Faculty Award [DST/INSPIRE/04/2014/002461]. We thank Arnab
Das for helpful comments.

\bibliography{refv_new}

\begin{thebibliography}{110}%
\makeatletter
\providecommand \@ifxundefined [1]{%
 \@ifx{#1\undefined}
}%
\providecommand \@ifnum [1]{%
 \ifnum #1\expandafter \@firstoftwo
 \else \expandafter \@secondoftwo
 \fi
}%
\providecommand \@ifx [1]{%
 \ifx #1\expandafter \@firstoftwo
 \else \expandafter \@secondoftwo
 \fi
}%
\providecommand \natexlab [1]{#1}%
\providecommand \enquote  [1]{``#1''}%
\providecommand \bibnamefont  [1]{#1}%
\providecommand \bibfnamefont [1]{#1}%
\providecommand \citenamefont [1]{#1}%
\providecommand \href@noop [0]{\@secondoftwo}%
\providecommand \href [0]{\begingroup \@sanitize@url \@href}%
\providecommand \@href[1]{\@@startlink{#1}\@@href}%
\providecommand \@@href[1]{\endgroup#1\@@endlink}%
\providecommand \@sanitize@url [0]{\catcode `\\12\catcode `\$12\catcode
  `\&12\catcode `\#12\catcode `\^12\catcode `\_12\catcode `\%12\relax}%
\providecommand \@@startlink[1]{}%
\providecommand \@@endlink[0]{}%
\providecommand \url  [0]{\begingroup\@sanitize@url \@url }%
\providecommand \@url [1]{\endgroup\@href {#1}{\urlprefix }}%
\providecommand \urlprefix  [0]{URL }%
\providecommand \Eprint [0]{\href }%
\providecommand \doibase [0]{https://doi.org/}%
\providecommand \selectlanguage [0]{\@gobble}%
\providecommand \bibinfo  [0]{\@secondoftwo}%
\providecommand \bibfield  [0]{\@secondoftwo}%
\providecommand \translation [1]{[#1]}%
\providecommand \BibitemOpen [0]{}%
\providecommand \bibitemStop [0]{}%
\providecommand \bibitemNoStop [0]{.\EOS\space}%
\providecommand \EOS [0]{\spacefactor3000\relax}%
\providecommand \BibitemShut  [1]{\csname bibitem#1\endcsname}%
\let\auto@bib@innerbib\@empty
\bibitem [{\citenamefont {Grifoni}\ and\ \citenamefont
  {Hänggi}(1998)}]{grifoni1998driven}%
  \BibitemOpen
  \bibfield  {author} {\bibinfo {author} {\bibfnamefont {M.}~\bibnamefont
  {Grifoni}}\ and\ \bibinfo {author} {\bibfnamefont {P.}~\bibnamefont
  {Hänggi}},\ }\bibfield  {title} {\bibinfo {title} {Driven quantum
  tunneling},\ }\href
  {https://doi.org/https://doi.org/10.1016/S0370-1573(98)00022-2} {\bibfield
  {journal} {\bibinfo  {journal} {Physics Reports}\ }\textbf {\bibinfo {volume}
  {304}},\ \bibinfo {pages} {229} (\bibinfo {year} {1998})}\BibitemShut
  {NoStop}%
\bibitem [{\citenamefont {D'Alessio}\ and\ \citenamefont
  {Rigol}(2014)}]{dalessio2014long}%
  \BibitemOpen
  \bibfield  {author} {\bibinfo {author} {\bibfnamefont {L.}~\bibnamefont
  {D'Alessio}}\ and\ \bibinfo {author} {\bibfnamefont {M.}~\bibnamefont
  {Rigol}},\ }\bibfield  {title} {\bibinfo {title} {Long-time behavior of
  isolated periodically driven interacting lattice systems},\ }\href
  {https://doi.org/10.1103/PhysRevX.4.041048} {\bibfield  {journal} {\bibinfo
  {journal} {Phys. Rev. X}\ }\textbf {\bibinfo {volume} {4}},\ \bibinfo {pages}
  {041048} (\bibinfo {year} {2014})}\BibitemShut {NoStop}%
\bibitem [{\citenamefont {Lazarides}\ \emph
  {et~al.}(2014{\natexlab{a}})\citenamefont {Lazarides}, \citenamefont {Das},\
  and\ \citenamefont {Moessner}}]{lazarides2014equilibrium}%
  \BibitemOpen
  \bibfield  {author} {\bibinfo {author} {\bibfnamefont {A.}~\bibnamefont
  {Lazarides}}, \bibinfo {author} {\bibfnamefont {A.}~\bibnamefont {Das}},\
  and\ \bibinfo {author} {\bibfnamefont {R.}~\bibnamefont {Moessner}},\
  }\bibfield  {title} {\bibinfo {title} {Equilibrium states of generic quantum
  systems subject to periodic driving},\ }\href
  {https://doi.org/10.1103/PhysRevE.90.012110} {\bibfield  {journal} {\bibinfo
  {journal} {Phys. Rev. E}\ }\textbf {\bibinfo {volume} {90}},\ \bibinfo
  {pages} {012110} (\bibinfo {year} {2014}{\natexlab{a}})}\BibitemShut
  {NoStop}%
\bibitem [{\citenamefont {Lazarides}\ \emph
  {et~al.}(2014{\natexlab{b}})\citenamefont {Lazarides}, \citenamefont {Das},\
  and\ \citenamefont {Moessner}}]{Lazarides2014periodic}%
  \BibitemOpen
  \bibfield  {author} {\bibinfo {author} {\bibfnamefont {A.}~\bibnamefont
  {Lazarides}}, \bibinfo {author} {\bibfnamefont {A.}~\bibnamefont {Das}},\
  and\ \bibinfo {author} {\bibfnamefont {R.}~\bibnamefont {Moessner}},\
  }\bibfield  {title} {\bibinfo {title} {Periodic thermodynamics of isolated
  quantum systems},\ }\href {https://doi.org/10.1103/PhysRevLett.112.150401}
  {\bibfield  {journal} {\bibinfo  {journal} {Phys. Rev. Lett.}\ }\textbf
  {\bibinfo {volume} {112}},\ \bibinfo {pages} {150401} (\bibinfo {year}
  {2014}{\natexlab{b}})}\BibitemShut {NoStop}%
\bibitem [{\citenamefont {Bukov}\ \emph {et~al.}(2015)\citenamefont {Bukov},
  \citenamefont {D'Alessio},\ and\ \citenamefont
  {Polkovnikov}}]{bukov2015universal}%
  \BibitemOpen
  \bibfield  {author} {\bibinfo {author} {\bibfnamefont {M.}~\bibnamefont
  {Bukov}}, \bibinfo {author} {\bibfnamefont {L.}~\bibnamefont {D'Alessio}},\
  and\ \bibinfo {author} {\bibfnamefont {A.}~\bibnamefont {Polkovnikov}},\
  }\bibfield  {title} {\bibinfo {title} {{Universal high-frequency behavior of
  periodically driven systems: from dynamical stabilization to Floquet
  engineering}},\ }\href {https://doi.org/10.1080/00018732.2015.1055918}
  {\bibfield  {journal} {\bibinfo  {journal} {Advances in Physics}\ }\textbf
  {\bibinfo {volume} {64}},\ \bibinfo {pages} {139} (\bibinfo {year}
  {2015})}\BibitemShut {NoStop}%
\bibitem [{\citenamefont {Eckardt}\ and\ \citenamefont
  {Anisimovas}(2015)}]{eckardt2015high}%
  \BibitemOpen
  \bibfield  {author} {\bibinfo {author} {\bibfnamefont {A.}~\bibnamefont
  {Eckardt}}\ and\ \bibinfo {author} {\bibfnamefont {E.}~\bibnamefont
  {Anisimovas}},\ }\bibfield  {title} {\bibinfo {title} {{High-frequency
  approximation for periodically driven quantum systems from a Floquet-space
  perspective}},\ }\href {https://doi.org/10.1088/1367-2630/17/9/093039}
  {\bibfield  {journal} {\bibinfo  {journal} {New Journal of Physics}\ }\textbf
  {\bibinfo {volume} {17}},\ \bibinfo {pages} {093039} (\bibinfo {year}
  {2015})}\BibitemShut {NoStop}%
\bibitem [{\citenamefont {Ponte}\ \emph
  {et~al.}(2015{\natexlab{a}})\citenamefont {Ponte}, \citenamefont {Chandran},
  \citenamefont {Papić},\ and\ \citenamefont
  {Abanin}}]{ponte2015periodically}%
  \BibitemOpen
  \bibfield  {author} {\bibinfo {author} {\bibfnamefont {P.}~\bibnamefont
  {Ponte}}, \bibinfo {author} {\bibfnamefont {A.}~\bibnamefont {Chandran}},
  \bibinfo {author} {\bibfnamefont {Z.}~\bibnamefont {Papić}},\ and\ \bibinfo
  {author} {\bibfnamefont {D.~A.}\ \bibnamefont {Abanin}},\ }\bibfield  {title}
  {\bibinfo {title} {Periodically driven ergodic and many-body localized
  quantum systems},\ }\href
  {https://doi.org/https://doi.org/10.1016/j.aop.2014.11.008} {\bibfield
  {journal} {\bibinfo  {journal} {Annals of Physics}\ }\textbf {\bibinfo
  {volume} {353}},\ \bibinfo {pages} {196} (\bibinfo {year}
  {2015}{\natexlab{a}})}\BibitemShut {NoStop}%
\bibitem [{\citenamefont {Eisert}\ \emph {et~al.}(2015)\citenamefont {Eisert},
  \citenamefont {Friesdorf},\ and\ \citenamefont
  {Gogolin}}]{eisert2015quantum}%
  \BibitemOpen
  \bibfield  {author} {\bibinfo {author} {\bibfnamefont {J.}~\bibnamefont
  {Eisert}}, \bibinfo {author} {\bibfnamefont {M.}~\bibnamefont {Friesdorf}},\
  and\ \bibinfo {author} {\bibfnamefont {C.}~\bibnamefont {Gogolin}},\
  }\bibfield  {title} {\bibinfo {title} {Quantum many-body systems out of
  equilibrium},\ }\href {https://doi.org/10.1038/nphys3215} {\bibfield
  {journal} {\bibinfo  {journal} {Nature Physics}\ }\textbf {\bibinfo {volume}
  {11}},\ \bibinfo {pages} {124} (\bibinfo {year} {2015})}\BibitemShut
  {NoStop}%
\bibitem [{\citenamefont {Khemani}\ \emph {et~al.}(2016)\citenamefont
  {Khemani}, \citenamefont {Lazarides}, \citenamefont {Moessner},\ and\
  \citenamefont {Sondhi}}]{khemani2016phase}%
  \BibitemOpen
  \bibfield  {author} {\bibinfo {author} {\bibfnamefont {V.}~\bibnamefont
  {Khemani}}, \bibinfo {author} {\bibfnamefont {A.}~\bibnamefont {Lazarides}},
  \bibinfo {author} {\bibfnamefont {R.}~\bibnamefont {Moessner}},\ and\
  \bibinfo {author} {\bibfnamefont {S.~L.}\ \bibnamefont {Sondhi}},\ }\bibfield
   {title} {\bibinfo {title} {Phase structure of driven quantum systems},\
  }\href {https://doi.org/10.1103/PhysRevLett.116.250401} {\bibfield  {journal}
  {\bibinfo  {journal} {Phys. Rev. Lett.}\ }\textbf {\bibinfo {volume} {116}},\
  \bibinfo {pages} {250401} (\bibinfo {year} {2016})}\BibitemShut {NoStop}%
\bibitem [{\citenamefont {Ho}\ \emph {et~al.}(2022)\citenamefont {Ho},
  \citenamefont {Mori}, \citenamefont {Abanin},\ and\ \citenamefont
  {Torre}}]{ho2022quantum}%
  \BibitemOpen
  \bibfield  {author} {\bibinfo {author} {\bibfnamefont {W.~W.}\ \bibnamefont
  {Ho}}, \bibinfo {author} {\bibfnamefont {T.}~\bibnamefont {Mori}}, \bibinfo
  {author} {\bibfnamefont {D.~A.}\ \bibnamefont {Abanin}},\ and\ \bibinfo
  {author} {\bibfnamefont {E.~G.~D.}\ \bibnamefont {Torre}},\ }\bibfield
  {title} {\bibinfo {title} {{Quantum and classical Floquet
  prethermalization}}\ }\href {https://doi.org/10.48550/arxiv.2212.00041}
  {10.48550/arxiv.2212.00041} (\bibinfo {year} {2022})\BibitemShut {NoStop}%
\bibitem [{\citenamefont {Casati}\ and\ \citenamefont
  {Ford}(1979)}]{casati1979stochastic}%
  \BibitemOpen
  \bibfield  {author} {\bibinfo {author} {\bibfnamefont {G.}~\bibnamefont
  {Casati}}\ and\ \bibinfo {author} {\bibfnamefont {J.}~\bibnamefont {Ford}},\
  }\bibfield  {title} {\bibinfo {title} {{Stochastic behavior in classical and
  quantum Hamiltonian systems}}\ }\textbf {\bibinfo {volume} {93}},\ \href
  {https://doi.org/10.1007/BFB0021732} {10.1007/BFB0021732} (\bibinfo {year}
  {1979})\BibitemShut {NoStop}%
\bibitem [{\citenamefont {Grempel}\ \emph {et~al.}(1984)\citenamefont
  {Grempel}, \citenamefont {Prange},\ and\ \citenamefont
  {Fishman}}]{grempel19841uantum}%
  \BibitemOpen
  \bibfield  {author} {\bibinfo {author} {\bibfnamefont {D.~R.}\ \bibnamefont
  {Grempel}}, \bibinfo {author} {\bibfnamefont {R.~E.}\ \bibnamefont
  {Prange}},\ and\ \bibinfo {author} {\bibfnamefont {S.}~\bibnamefont
  {Fishman}},\ }\bibfield  {title} {\bibinfo {title} {Quantum dynamics of a
  nonintegrable system},\ }\href {https://doi.org/10.1103/PhysRevA.29.1639}
  {\bibfield  {journal} {\bibinfo  {journal} {Phys. Rev. A}\ }\textbf {\bibinfo
  {volume} {29}},\ \bibinfo {pages} {1639} (\bibinfo {year}
  {1984})}\BibitemShut {NoStop}%
\bibitem [{\citenamefont {Lemari\'e}\ \emph {et~al.}(2009)\citenamefont
  {Lemari\'e}, \citenamefont {Chab\'e}, \citenamefont {Szriftgiser},
  \citenamefont {Garreau}, \citenamefont {Gr\'emaud},\ and\ \citenamefont
  {Delande}}]{Lemarie2009observation}%
  \BibitemOpen
  \bibfield  {author} {\bibinfo {author} {\bibfnamefont {G.}~\bibnamefont
  {Lemari\'e}}, \bibinfo {author} {\bibfnamefont {J.}~\bibnamefont {Chab\'e}},
  \bibinfo {author} {\bibfnamefont {P.}~\bibnamefont {Szriftgiser}}, \bibinfo
  {author} {\bibfnamefont {J.~C.}\ \bibnamefont {Garreau}}, \bibinfo {author}
  {\bibfnamefont {B.}~\bibnamefont {Gr\'emaud}},\ and\ \bibinfo {author}
  {\bibfnamefont {D.}~\bibnamefont {Delande}},\ }\bibfield  {title} {\bibinfo
  {title} {Observation of the anderson metal-insulator transition with atomic
  matter waves: Theory and experiment},\ }\href
  {https://doi.org/10.1103/PhysRevA.80.043626} {\bibfield  {journal} {\bibinfo
  {journal} {Phys. Rev. A}\ }\textbf {\bibinfo {volume} {80}},\ \bibinfo
  {pages} {043626} (\bibinfo {year} {2009})}\BibitemShut {NoStop}%
\bibitem [{\citenamefont {Das}(2010)}]{das2010exotic}%
  \BibitemOpen
  \bibfield  {author} {\bibinfo {author} {\bibfnamefont {A.}~\bibnamefont
  {Das}},\ }\bibfield  {title} {\bibinfo {title} {Exotic freezing of response
  in a quantum many-body system},\ }\href
  {https://doi.org/10.1103/PhysRevB.82.172402} {\bibfield  {journal} {\bibinfo
  {journal} {Phys. Rev. B}\ }\textbf {\bibinfo {volume} {82}},\ \bibinfo
  {pages} {172402} (\bibinfo {year} {2010})}\BibitemShut {NoStop}%
\bibitem [{\citenamefont {Roy}\ and\ \citenamefont {Das}(2015)}]{roy2015fate}%
  \BibitemOpen
  \bibfield  {author} {\bibinfo {author} {\bibfnamefont {A.}~\bibnamefont
  {Roy}}\ and\ \bibinfo {author} {\bibfnamefont {A.}~\bibnamefont {Das}},\
  }\bibfield  {title} {\bibinfo {title} {Fate of dynamical many-body
  localization in the presence of disorder},\ }\href
  {https://doi.org/10.1103/PhysRevB.91.121106} {\bibfield  {journal} {\bibinfo
  {journal} {Phys. Rev. B}\ }\textbf {\bibinfo {volume} {91}},\ \bibinfo
  {pages} {121106} (\bibinfo {year} {2015})}\BibitemShut {NoStop}%
\bibitem [{\citenamefont {Haldar}\ and\ \citenamefont
  {Das}(2017)}]{haldar2017dynamical}%
  \BibitemOpen
  \bibfield  {author} {\bibinfo {author} {\bibfnamefont {A.}~\bibnamefont
  {Haldar}}\ and\ \bibinfo {author} {\bibfnamefont {A.}~\bibnamefont {Das}},\
  }\bibfield  {title} {\bibinfo {title} {Dynamical many-body localization and
  delocalization in periodically driven closed quantum systems},\ }\href
  {https://doi.org/https://doi.org/10.1002/andp.201600333} {\bibfield
  {journal} {\bibinfo  {journal} {Annalen der Physik}\ }\textbf {\bibinfo
  {volume} {529}},\ \bibinfo {pages} {1600333} (\bibinfo {year}
  {2017})}\BibitemShut {NoStop}%
\bibitem [{\citenamefont {Haldar}\ \emph {et~al.}(2018)\citenamefont {Haldar},
  \citenamefont {Moessner},\ and\ \citenamefont {Das}}]{haldar2018onset}%
  \BibitemOpen
  \bibfield  {author} {\bibinfo {author} {\bibfnamefont {A.}~\bibnamefont
  {Haldar}}, \bibinfo {author} {\bibfnamefont {R.}~\bibnamefont {Moessner}},\
  and\ \bibinfo {author} {\bibfnamefont {A.}~\bibnamefont {Das}},\ }\bibfield
  {title} {\bibinfo {title} {{Onset of Floquet thermalization}},\ }\href
  {https://doi.org/10.1103/PhysRevB.97.245122} {\bibfield  {journal} {\bibinfo
  {journal} {Phys. Rev. B}\ }\textbf {\bibinfo {volume} {97}},\ \bibinfo
  {pages} {245122} (\bibinfo {year} {2018})}\BibitemShut {NoStop}%
\bibitem [{\citenamefont {Haldar}\ \emph {et~al.}(2021)\citenamefont {Haldar},
  \citenamefont {Sen}, \citenamefont {Moessner},\ and\ \citenamefont
  {Das}}]{haldar2021dynamical}%
  \BibitemOpen
  \bibfield  {author} {\bibinfo {author} {\bibfnamefont {A.}~\bibnamefont
  {Haldar}}, \bibinfo {author} {\bibfnamefont {D.}~\bibnamefont {Sen}},
  \bibinfo {author} {\bibfnamefont {R.}~\bibnamefont {Moessner}},\ and\
  \bibinfo {author} {\bibfnamefont {A.}~\bibnamefont {Das}},\ }\bibfield
  {title} {\bibinfo {title} {{Dynamical Freezing and Scar Points in Strongly
  Driven Floquet Matter: Resonance vs Emergent Conservation Laws}},\ }\href
  {https://doi.org/10.1103/PhysRevX.11.021008} {\bibfield  {journal} {\bibinfo
  {journal} {Phys. Rev. X}\ }\textbf {\bibinfo {volume} {11}},\ \bibinfo
  {pages} {021008} (\bibinfo {year} {2021})}\BibitemShut {NoStop}%
\bibitem [{\citenamefont {Haldar}\ and\ \citenamefont
  {Das}(2022)}]{Haldar2022statistical}%
  \BibitemOpen
  \bibfield  {author} {\bibinfo {author} {\bibfnamefont {A.}~\bibnamefont
  {Haldar}}\ and\ \bibinfo {author} {\bibfnamefont {A.}~\bibnamefont {Das}},\
  }\bibfield  {title} {\bibinfo {title} {{Statistical mechanics of Floquet
  quantum matter: exact and emergent conservation laws}},\ }\href
  {https://doi.org/10.1088/1361-648X/ac03d2} {\bibfield  {journal} {\bibinfo
  {journal} {Journal of Physics: Condensed Matter}\ }\textbf {\bibinfo {volume}
  {34}},\ \bibinfo {pages} {234001} (\bibinfo {year} {2022})}\BibitemShut
  {NoStop}%
\bibitem [{\citenamefont {Kitagawa}\ \emph {et~al.}(2010)\citenamefont
  {Kitagawa}, \citenamefont {Berg}, \citenamefont {Rudner},\ and\ \citenamefont
  {Demler}}]{kitagawa2010topological}%
  \BibitemOpen
  \bibfield  {author} {\bibinfo {author} {\bibfnamefont {T.}~\bibnamefont
  {Kitagawa}}, \bibinfo {author} {\bibfnamefont {E.}~\bibnamefont {Berg}},
  \bibinfo {author} {\bibfnamefont {M.}~\bibnamefont {Rudner}},\ and\ \bibinfo
  {author} {\bibfnamefont {E.}~\bibnamefont {Demler}},\ }\bibfield  {title}
  {\bibinfo {title} {Topological characterization of periodically driven
  quantum systems},\ }\href {https://doi.org/10.1103/PhysRevB.82.235114}
  {\bibfield  {journal} {\bibinfo  {journal} {Phys. Rev. B}\ }\textbf {\bibinfo
  {volume} {82}},\ \bibinfo {pages} {235114} (\bibinfo {year}
  {2010})}\BibitemShut {NoStop}%
\bibitem [{\citenamefont {Lindner}\ \emph {et~al.}(2011)\citenamefont
  {Lindner}, \citenamefont {Refael},\ and\ \citenamefont
  {Galitski}}]{lindner2011floquet}%
  \BibitemOpen
  \bibfield  {author} {\bibinfo {author} {\bibfnamefont {N.~H.}\ \bibnamefont
  {Lindner}}, \bibinfo {author} {\bibfnamefont {G.}~\bibnamefont {Refael}},\
  and\ \bibinfo {author} {\bibfnamefont {V.}~\bibnamefont {Galitski}},\
  }\bibfield  {title} {\bibinfo {title} {{Floquet topological insulator in
  semiconductor quantum wells}},\ }\href {https://doi.org/10.1038/nphys1926}
  {\bibfield  {journal} {\bibinfo  {journal} {Nature Physics}\ }\textbf
  {\bibinfo {volume} {7}},\ \bibinfo {pages} {490} (\bibinfo {year}
  {2011})}\BibitemShut {NoStop}%
\bibitem [{\citenamefont {Nathan}\ \emph {et~al.}(2019)\citenamefont {Nathan},
  \citenamefont {Abanin}, \citenamefont {Berg}, \citenamefont {Lindner},\ and\
  \citenamefont {Rudner}}]{nathananomalous2019}%
  \BibitemOpen
  \bibfield  {author} {\bibinfo {author} {\bibfnamefont {F.}~\bibnamefont
  {Nathan}}, \bibinfo {author} {\bibfnamefont {D.}~\bibnamefont {Abanin}},
  \bibinfo {author} {\bibfnamefont {E.}~\bibnamefont {Berg}}, \bibinfo {author}
  {\bibfnamefont {N.~H.}\ \bibnamefont {Lindner}},\ and\ \bibinfo {author}
  {\bibfnamefont {M.~S.}\ \bibnamefont {Rudner}},\ }\bibfield  {title}
  {\bibinfo {title} {{Anomalous Floquet insulators}},\ }\href
  {https://doi.org/10.1103/PhysRevB.99.195133} {\bibfield  {journal} {\bibinfo
  {journal} {Phys. Rev. B}\ }\textbf {\bibinfo {volume} {99}},\ \bibinfo
  {pages} {195133} (\bibinfo {year} {2019})}\BibitemShut {NoStop}%
\bibitem [{\citenamefont {Abanin}\ \emph {et~al.}(2015)\citenamefont {Abanin},
  \citenamefont {De~Roeck},\ and\ \citenamefont
  {Huveneers}}]{abanin2015exponentially}%
  \BibitemOpen
  \bibfield  {author} {\bibinfo {author} {\bibfnamefont {D.~A.}\ \bibnamefont
  {Abanin}}, \bibinfo {author} {\bibfnamefont {W.}~\bibnamefont {De~Roeck}},\
  and\ \bibinfo {author} {\bibfnamefont {F.}~\bibnamefont {Huveneers}},\
  }\bibfield  {title} {\bibinfo {title} {Exponentially slow heating in
  periodically driven many-body systems},\ }\href
  {https://doi.org/10.1103/PhysRevLett.115.256803} {\bibfield  {journal}
  {\bibinfo  {journal} {Phys. Rev. Lett.}\ }\textbf {\bibinfo {volume} {115}},\
  \bibinfo {pages} {256803} (\bibinfo {year} {2015})}\BibitemShut {NoStop}%
\bibitem [{\citenamefont {Mori}\ \emph {et~al.}(2016)\citenamefont {Mori},
  \citenamefont {Kuwahara},\ and\ \citenamefont {Saito}}]{mori2016rigorous}%
  \BibitemOpen
  \bibfield  {author} {\bibinfo {author} {\bibfnamefont {T.}~\bibnamefont
  {Mori}}, \bibinfo {author} {\bibfnamefont {T.}~\bibnamefont {Kuwahara}},\
  and\ \bibinfo {author} {\bibfnamefont {K.}~\bibnamefont {Saito}},\ }\bibfield
   {title} {\bibinfo {title} {Rigorous bound on energy absorption and generic
  relaxation in periodically driven quantum systems},\ }\href
  {https://doi.org/10.1103/PhysRevLett.116.120401} {\bibfield  {journal}
  {\bibinfo  {journal} {Phys. Rev. Lett.}\ }\textbf {\bibinfo {volume} {116}},\
  \bibinfo {pages} {120401} (\bibinfo {year} {2016})}\BibitemShut {NoStop}%
\bibitem [{\citenamefont {Abanin}\ \emph
  {et~al.}(2017{\natexlab{a}})\citenamefont {Abanin}, \citenamefont {De~Roeck},
  \citenamefont {Ho},\ and\ \citenamefont {Huveneers}}]{abanin2017effective}%
  \BibitemOpen
  \bibfield  {author} {\bibinfo {author} {\bibfnamefont {D.~A.}\ \bibnamefont
  {Abanin}}, \bibinfo {author} {\bibfnamefont {W.}~\bibnamefont {De~Roeck}},
  \bibinfo {author} {\bibfnamefont {W.~W.}\ \bibnamefont {Ho}},\ and\ \bibinfo
  {author} {\bibfnamefont {F.}~\bibnamefont {Huveneers}},\ }\bibfield  {title}
  {\bibinfo {title} {{Effective Hamiltonians, prethermalization, and slow
  energy absorption in periodically driven many-body systems}},\ }\href
  {https://doi.org/10.1103/PhysRevB.95.014112} {\bibfield  {journal} {\bibinfo
  {journal} {Phys. Rev. B}\ }\textbf {\bibinfo {volume} {95}},\ \bibinfo
  {pages} {014112} (\bibinfo {year} {2017}{\natexlab{a}})}\BibitemShut
  {NoStop}%
\bibitem [{\citenamefont {Abanin}\ \emph
  {et~al.}(2017{\natexlab{b}})\citenamefont {Abanin}, \citenamefont {De~Roeck},
  \citenamefont {Ho},\ and\ \citenamefont {Huveneers}}]{abanin2017rigorous}%
  \BibitemOpen
  \bibfield  {author} {\bibinfo {author} {\bibfnamefont {D.}~\bibnamefont
  {Abanin}}, \bibinfo {author} {\bibfnamefont {W.}~\bibnamefont {De~Roeck}},
  \bibinfo {author} {\bibfnamefont {W.~W.}\ \bibnamefont {Ho}},\ and\ \bibinfo
  {author} {\bibfnamefont {F.}~\bibnamefont {Huveneers}},\ }\bibfield  {title}
  {\bibinfo {title} {A rigorous theory of many-body prethermalization for
  periodically driven and closed quantum systems},\ }\href
  {https://doi.org/10.1007/s00220-017-2930-x} {\bibfield  {journal} {\bibinfo
  {journal} {Communications in Mathematical Physics}\ }\textbf {\bibinfo
  {volume} {354}},\ \bibinfo {pages} {809} (\bibinfo {year}
  {2017}{\natexlab{b}})}\BibitemShut {NoStop}%
\bibitem [{\citenamefont {Weidinger}\ and\ \citenamefont
  {Knap}(2017)}]{weidinger2017floquet}%
  \BibitemOpen
  \bibfield  {author} {\bibinfo {author} {\bibfnamefont {S.~A.}\ \bibnamefont
  {Weidinger}}\ and\ \bibinfo {author} {\bibfnamefont {M.}~\bibnamefont
  {Knap}},\ }\bibfield  {title} {\bibinfo {title} {{Floquet prethermalization
  and regimes of heating in a periodically driven, interacting quantum
  system}},\ }\href {https://doi.org/10.1038/srep45382} {\bibfield  {journal}
  {\bibinfo  {journal} {Scientific Reports}\ }\textbf {\bibinfo {volume} {7}},\
  \bibinfo {pages} {45382} (\bibinfo {year} {2017})}\BibitemShut {NoStop}%
\bibitem [{\citenamefont {Beatrez}\ \emph {et~al.}(2021)\citenamefont
  {Beatrez}, \citenamefont {Janes}, \citenamefont {Akkiraju}, \citenamefont
  {Pillai}, \citenamefont {Oddo}, \citenamefont {Reshetikhin}, \citenamefont
  {Druga}, \citenamefont {McAllister}, \citenamefont {Elo}, \citenamefont
  {Gilbert}, \citenamefont {Suter},\ and\ \citenamefont
  {Ajoy}}]{beatrez2021floquet}%
  \BibitemOpen
  \bibfield  {author} {\bibinfo {author} {\bibfnamefont {W.}~\bibnamefont
  {Beatrez}}, \bibinfo {author} {\bibfnamefont {O.}~\bibnamefont {Janes}},
  \bibinfo {author} {\bibfnamefont {A.}~\bibnamefont {Akkiraju}}, \bibinfo
  {author} {\bibfnamefont {A.}~\bibnamefont {Pillai}}, \bibinfo {author}
  {\bibfnamefont {A.}~\bibnamefont {Oddo}}, \bibinfo {author} {\bibfnamefont
  {P.}~\bibnamefont {Reshetikhin}}, \bibinfo {author} {\bibfnamefont
  {E.}~\bibnamefont {Druga}}, \bibinfo {author} {\bibfnamefont
  {M.}~\bibnamefont {McAllister}}, \bibinfo {author} {\bibfnamefont
  {M.}~\bibnamefont {Elo}}, \bibinfo {author} {\bibfnamefont {B.}~\bibnamefont
  {Gilbert}}, \bibinfo {author} {\bibfnamefont {D.}~\bibnamefont {Suter}},\
  and\ \bibinfo {author} {\bibfnamefont {A.}~\bibnamefont {Ajoy}},\ }\bibfield
  {title} {\bibinfo {title} {{Floquet Prethermalization with Lifetime Exceeding
  90 s in a Bulk Hyperpolarized Solid}},\ }\href
  {https://doi.org/10.1103/PhysRevLett.127.170603} {\bibfield  {journal}
  {\bibinfo  {journal} {Phys. Rev. Lett.}\ }\textbf {\bibinfo {volume} {127}},\
  \bibinfo {pages} {170603} (\bibinfo {year} {2021})}\BibitemShut {NoStop}%
\bibitem [{\citenamefont {Peng}\ \emph {et~al.}(2021)\citenamefont {Peng},
  \citenamefont {Yin}, \citenamefont {Huang}, \citenamefont {Ramanathan},\ and\
  \citenamefont {Cappellaro}}]{peng2021floquet}%
  \BibitemOpen
  \bibfield  {author} {\bibinfo {author} {\bibfnamefont {P.}~\bibnamefont
  {Peng}}, \bibinfo {author} {\bibfnamefont {C.}~\bibnamefont {Yin}}, \bibinfo
  {author} {\bibfnamefont {X.}~\bibnamefont {Huang}}, \bibinfo {author}
  {\bibfnamefont {C.}~\bibnamefont {Ramanathan}},\ and\ \bibinfo {author}
  {\bibfnamefont {P.}~\bibnamefont {Cappellaro}},\ }\bibfield  {title}
  {\bibinfo {title} {{Floquet prethermalization in dipolar spin chains}},\
  }\href {https://doi.org/10.1038/s41567-020-01120-z} {\bibfield  {journal}
  {\bibinfo  {journal} {Nature Physics}\ }\textbf {\bibinfo {volume} {17}},\
  \bibinfo {pages} {444} (\bibinfo {year} {2021})}\BibitemShut {NoStop}%
\bibitem [{\citenamefont {Fleckenstein}\ and\ \citenamefont
  {Bukov}(2021{\natexlab{a}})}]{fleckenstein2021prethermalization}%
  \BibitemOpen
  \bibfield  {author} {\bibinfo {author} {\bibfnamefont {C.}~\bibnamefont
  {Fleckenstein}}\ and\ \bibinfo {author} {\bibfnamefont {M.}~\bibnamefont
  {Bukov}},\ }\bibfield  {title} {\bibinfo {title} {Prethermalization and
  thermalization in periodically driven many-body systems away from the
  high-frequency limit},\ }\href {https://doi.org/10.1103/PhysRevB.103.L140302}
  {\bibfield  {journal} {\bibinfo  {journal} {Phys. Rev. B}\ }\textbf {\bibinfo
  {volume} {103}},\ \bibinfo {pages} {L140302} (\bibinfo {year}
  {2021}{\natexlab{a}})}\BibitemShut {NoStop}%
\bibitem [{\citenamefont {Fleckenstein}\ and\ \citenamefont
  {Bukov}(2021{\natexlab{b}})}]{fleckenstein2021prethermalization_1}%
  \BibitemOpen
  \bibfield  {author} {\bibinfo {author} {\bibfnamefont {C.}~\bibnamefont
  {Fleckenstein}}\ and\ \bibinfo {author} {\bibfnamefont {M.}~\bibnamefont
  {Bukov}},\ }\bibfield  {title} {\bibinfo {title} {Thermalization and
  prethermalization in periodically kicked quantum spin chains},\ }\href
  {https://doi.org/10.1103/PhysRevB.103.144307} {\bibfield  {journal} {\bibinfo
   {journal} {Phys. Rev. B}\ }\textbf {\bibinfo {volume} {103}},\ \bibinfo
  {pages} {144307} (\bibinfo {year} {2021}{\natexlab{b}})}\BibitemShut
  {NoStop}%
\bibitem [{\citenamefont {Morningstar}\ \emph {et~al.}(2022)\citenamefont
  {Morningstar}, \citenamefont {Hauru}, \citenamefont {Beall}, \citenamefont
  {Ganahl}, \citenamefont {Lewis}, \citenamefont {Khemani},\ and\ \citenamefont
  {Vidal}}]{morningstar2022simulation}%
  \BibitemOpen
  \bibfield  {author} {\bibinfo {author} {\bibfnamefont {A.}~\bibnamefont
  {Morningstar}}, \bibinfo {author} {\bibfnamefont {M.}~\bibnamefont {Hauru}},
  \bibinfo {author} {\bibfnamefont {J.}~\bibnamefont {Beall}}, \bibinfo
  {author} {\bibfnamefont {M.}~\bibnamefont {Ganahl}}, \bibinfo {author}
  {\bibfnamefont {A.~G.}\ \bibnamefont {Lewis}}, \bibinfo {author}
  {\bibfnamefont {V.}~\bibnamefont {Khemani}},\ and\ \bibinfo {author}
  {\bibfnamefont {G.}~\bibnamefont {Vidal}},\ }\bibfield  {title} {\bibinfo
  {title} {{Simulation of Quantum Many-Body Dynamics with Tensor Processing
  Units: Floquet Prethermalization}},\ }\href
  {https://doi.org/10.1103/PRXQuantum.3.020331} {\bibfield  {journal} {\bibinfo
   {journal} {PRX Quantum}\ }\textbf {\bibinfo {volume} {3}},\ \bibinfo {pages}
  {020331} (\bibinfo {year} {2022})}\BibitemShut {NoStop}%
\bibitem [{\citenamefont {Else}\ \emph {et~al.}(2016)\citenamefont {Else},
  \citenamefont {Bauer},\ and\ \citenamefont {Nayak}}]{else2016floquet}%
  \BibitemOpen
  \bibfield  {author} {\bibinfo {author} {\bibfnamefont {D.~V.}\ \bibnamefont
  {Else}}, \bibinfo {author} {\bibfnamefont {B.}~\bibnamefont {Bauer}},\ and\
  \bibinfo {author} {\bibfnamefont {C.}~\bibnamefont {Nayak}},\ }\bibfield
  {title} {\bibinfo {title} {{Floquet Time Crystals}},\ }\href
  {https://doi.org/10.1103/PhysRevLett.117.090402} {\bibfield  {journal}
  {\bibinfo  {journal} {Phys. Rev. Lett.}\ }\textbf {\bibinfo {volume} {117}},\
  \bibinfo {pages} {090402} (\bibinfo {year} {2016})}\BibitemShut {NoStop}%
\bibitem [{\citenamefont {Yao}\ \emph {et~al.}(2017)\citenamefont {Yao},
  \citenamefont {Potter}, \citenamefont {Potirniche},\ and\ \citenamefont
  {Vishwanath}}]{yao2017discrete}%
  \BibitemOpen
  \bibfield  {author} {\bibinfo {author} {\bibfnamefont {N.~Y.}\ \bibnamefont
  {Yao}}, \bibinfo {author} {\bibfnamefont {A.~C.}\ \bibnamefont {Potter}},
  \bibinfo {author} {\bibfnamefont {I.-D.}\ \bibnamefont {Potirniche}},\ and\
  \bibinfo {author} {\bibfnamefont {A.}~\bibnamefont {Vishwanath}},\ }\bibfield
   {title} {\bibinfo {title} {Discrete time crystals: Rigidity, criticality,
  and realizations},\ }\href {https://doi.org/10.1103/PhysRevLett.118.030401}
  {\bibfield  {journal} {\bibinfo  {journal} {Phys. Rev. Lett.}\ }\textbf
  {\bibinfo {volume} {118}},\ \bibinfo {pages} {030401} (\bibinfo {year}
  {2017})}\BibitemShut {NoStop}%
\bibitem [{\citenamefont {Zhang}\ \emph {et~al.}(2017)\citenamefont {Zhang},
  \citenamefont {Hess}, \citenamefont {Kyprianidis}, \citenamefont {Becker},
  \citenamefont {Lee}, \citenamefont {Smith}, \citenamefont {Pagano},
  \citenamefont {Potirniche}, \citenamefont {Potter}, \citenamefont
  {Vishwanath}, \citenamefont {Yao},\ and\ \citenamefont
  {Monroe}}]{zhang2017observation}%
  \BibitemOpen
  \bibfield  {author} {\bibinfo {author} {\bibfnamefont {J.}~\bibnamefont
  {Zhang}}, \bibinfo {author} {\bibfnamefont {P.~W.}\ \bibnamefont {Hess}},
  \bibinfo {author} {\bibfnamefont {A.}~\bibnamefont {Kyprianidis}}, \bibinfo
  {author} {\bibfnamefont {P.}~\bibnamefont {Becker}}, \bibinfo {author}
  {\bibfnamefont {A.}~\bibnamefont {Lee}}, \bibinfo {author} {\bibfnamefont
  {J.}~\bibnamefont {Smith}}, \bibinfo {author} {\bibfnamefont
  {G.}~\bibnamefont {Pagano}}, \bibinfo {author} {\bibfnamefont {I.-D.}\
  \bibnamefont {Potirniche}}, \bibinfo {author} {\bibfnamefont {A.~C.}\
  \bibnamefont {Potter}}, \bibinfo {author} {\bibfnamefont {A.}~\bibnamefont
  {Vishwanath}}, \bibinfo {author} {\bibfnamefont {N.~Y.}\ \bibnamefont
  {Yao}},\ and\ \bibinfo {author} {\bibfnamefont {C.}~\bibnamefont {Monroe}},\
  }\bibfield  {title} {\bibinfo {title} {Observation of a discrete time
  crystal},\ }\href {https://doi.org/10.1038/nature21413} {\bibfield  {journal}
  {\bibinfo  {journal} {Nature}\ }\textbf {\bibinfo {volume} {543}},\ \bibinfo
  {pages} {217} (\bibinfo {year} {2017})}\BibitemShut {NoStop}%
\bibitem [{\citenamefont {Huang}\ \emph {et~al.}(2018)\citenamefont {Huang},
  \citenamefont {Wu},\ and\ \citenamefont {Liu}}]{huang2018clean}%
  \BibitemOpen
  \bibfield  {author} {\bibinfo {author} {\bibfnamefont {B.}~\bibnamefont
  {Huang}}, \bibinfo {author} {\bibfnamefont {Y.-H.}\ \bibnamefont {Wu}},\ and\
  \bibinfo {author} {\bibfnamefont {W.~V.}\ \bibnamefont {Liu}},\ }\bibfield
  {title} {\bibinfo {title} {{Clean Floquet Time Crystals: Models and
  Realizations in Cold Atoms}},\ }\href
  {https://doi.org/10.1103/PhysRevLett.120.110603} {\bibfield  {journal}
  {\bibinfo  {journal} {Phys. Rev. Lett.}\ }\textbf {\bibinfo {volume} {120}},\
  \bibinfo {pages} {110603} (\bibinfo {year} {2018})}\BibitemShut {NoStop}%
\bibitem [{\citenamefont {Khemani}\ \emph {et~al.}(2019)\citenamefont
  {Khemani}, \citenamefont {Moessner},\ and\ \citenamefont
  {Sondhi}}]{khemani2019brief}%
  \BibitemOpen
  \bibfield  {author} {\bibinfo {author} {\bibfnamefont {V.}~\bibnamefont
  {Khemani}}, \bibinfo {author} {\bibfnamefont {R.}~\bibnamefont {Moessner}},\
  and\ \bibinfo {author} {\bibfnamefont {S.~L.}\ \bibnamefont {Sondhi}},\
  }\bibfield  {title} {\bibinfo {title} {A brief history of time crystals}\
  }\href {https://doi.org/10.48550/arxiv.1910.10745}
  {10.48550/arxiv.1910.10745} (\bibinfo {year} {2019})\BibitemShut {NoStop}%
\bibitem [{\citenamefont {Lazarides}\ \emph {et~al.}(2015)\citenamefont
  {Lazarides}, \citenamefont {Das},\ and\ \citenamefont
  {Moessner}}]{lazarides2015fate}%
  \BibitemOpen
  \bibfield  {author} {\bibinfo {author} {\bibfnamefont {A.}~\bibnamefont
  {Lazarides}}, \bibinfo {author} {\bibfnamefont {A.}~\bibnamefont {Das}},\
  and\ \bibinfo {author} {\bibfnamefont {R.}~\bibnamefont {Moessner}},\
  }\bibfield  {title} {\bibinfo {title} {Fate of many-body localization under
  periodic driving},\ }\href {https://doi.org/10.1103/PhysRevLett.115.030402}
  {\bibfield  {journal} {\bibinfo  {journal} {Phys. Rev. Lett.}\ }\textbf
  {\bibinfo {volume} {115}},\ \bibinfo {pages} {030402} (\bibinfo {year}
  {2015})}\BibitemShut {NoStop}%
\bibitem [{\citenamefont {Ponte}\ \emph
  {et~al.}(2015{\natexlab{b}})\citenamefont {Ponte}, \citenamefont {Papi{\'c}},
  \citenamefont {Huveneers},\ and\ \citenamefont {Abanin}}]{ponte2015many}%
  \BibitemOpen
  \bibfield  {author} {\bibinfo {author} {\bibfnamefont {P.}~\bibnamefont
  {Ponte}}, \bibinfo {author} {\bibfnamefont {Z.}~\bibnamefont {Papi{\'c}}},
  \bibinfo {author} {\bibfnamefont {F.}~\bibnamefont {Huveneers}},\ and\
  \bibinfo {author} {\bibfnamefont {D.~A.}\ \bibnamefont {Abanin}},\ }\bibfield
   {title} {\bibinfo {title} {Many-body localization in periodically driven
  systems},\ }\href {https://doi.org/10.1103/PhysRevLett.114.140401} {\bibfield
   {journal} {\bibinfo  {journal} {Phys. Rev. Lett.}\ }\textbf {\bibinfo
  {volume} {114}},\ \bibinfo {pages} {140401} (\bibinfo {year}
  {2015}{\natexlab{b}})}\BibitemShut {NoStop}%
\bibitem [{\citenamefont {Abanin}\ \emph {et~al.}(2016)\citenamefont {Abanin},
  \citenamefont {{De Roeck}},\ and\ \citenamefont
  {Huveneers}}]{abanin2016theory}%
  \BibitemOpen
  \bibfield  {author} {\bibinfo {author} {\bibfnamefont {D.~A.}\ \bibnamefont
  {Abanin}}, \bibinfo {author} {\bibfnamefont {W.}~\bibnamefont {{De Roeck}}},\
  and\ \bibinfo {author} {\bibfnamefont {F.}~\bibnamefont {Huveneers}},\
  }\bibfield  {title} {\bibinfo {title} {Theory of many-body localization in
  periodically driven systems},\ }\href
  {https://doi.org/https://doi.org/10.1016/j.aop.2016.03.010} {\bibfield
  {journal} {\bibinfo  {journal} {Annals of Physics}\ }\textbf {\bibinfo
  {volume} {372}},\ \bibinfo {pages} {1} (\bibinfo {year} {2016})}\BibitemShut
  {NoStop}%
\bibitem [{\citenamefont {Dunlap}\ and\ \citenamefont
  {Kenkre}(1986)}]{dunlap1986dynamic}%
  \BibitemOpen
  \bibfield  {author} {\bibinfo {author} {\bibfnamefont {D.~H.}\ \bibnamefont
  {Dunlap}}\ and\ \bibinfo {author} {\bibfnamefont {V.~M.}\ \bibnamefont
  {Kenkre}},\ }\bibfield  {title} {\bibinfo {title} {Dynamic localization of a
  charged particle moving under the influence of an electric field},\ }\href
  {https://doi.org/10.1103/PhysRevB.34.3625} {\bibfield  {journal} {\bibinfo
  {journal} {Phys. Rev. B}\ }\textbf {\bibinfo {volume} {34}},\ \bibinfo
  {pages} {3625} (\bibinfo {year} {1986})}\BibitemShut {NoStop}%
\bibitem [{\citenamefont {Dunlap}\ and\ \citenamefont
  {Kenkre}(1988)}]{dunlap1988dynamic}%
  \BibitemOpen
  \bibfield  {author} {\bibinfo {author} {\bibfnamefont {D.}~\bibnamefont
  {Dunlap}}\ and\ \bibinfo {author} {\bibfnamefont {V.}~\bibnamefont
  {Kenkre}},\ }\bibfield  {title} {\bibinfo {title} {{Dynamic localization of a
  particle in an electric field viewed in momentum space: Connection with Bloch
  oscillations}},\ }\href
  {https://doi.org/https://doi.org/10.1016/0375-9601(88)90213-7} {\bibfield
  {journal} {\bibinfo  {journal} {Physics Letters A}\ }\textbf {\bibinfo
  {volume} {127}},\ \bibinfo {pages} {438} (\bibinfo {year}
  {1988})}\BibitemShut {NoStop}%
\bibitem [{\citenamefont {Lignier}\ \emph {et~al.}(2007)\citenamefont
  {Lignier}, \citenamefont {Sias}, \citenamefont {Ciampini}, \citenamefont
  {Singh}, \citenamefont {Zenesini}, \citenamefont {Morsch},\ and\
  \citenamefont {Arimondo}}]{lignier2007dynamical}%
  \BibitemOpen
  \bibfield  {author} {\bibinfo {author} {\bibfnamefont {H.}~\bibnamefont
  {Lignier}}, \bibinfo {author} {\bibfnamefont {C.}~\bibnamefont {Sias}},
  \bibinfo {author} {\bibfnamefont {D.}~\bibnamefont {Ciampini}}, \bibinfo
  {author} {\bibfnamefont {Y.}~\bibnamefont {Singh}}, \bibinfo {author}
  {\bibfnamefont {A.}~\bibnamefont {Zenesini}}, \bibinfo {author}
  {\bibfnamefont {O.}~\bibnamefont {Morsch}},\ and\ \bibinfo {author}
  {\bibfnamefont {E.}~\bibnamefont {Arimondo}},\ }\bibfield  {title} {\bibinfo
  {title} {Dynamical control of matter-wave tunneling in periodic potentials},\
  }\href {https://doi.org/10.1103/PhysRevLett.99.220403} {\bibfield  {journal}
  {\bibinfo  {journal} {Phys. Rev. Lett.}\ }\textbf {\bibinfo {volume} {99}},\
  \bibinfo {pages} {220403} (\bibinfo {year} {2007})}\BibitemShut {NoStop}%
\bibitem [{\citenamefont {Eckardt}\ \emph {et~al.}(2009)\citenamefont
  {Eckardt}, \citenamefont {Holthaus}, \citenamefont {Lignier}, \citenamefont
  {Zenesini}, \citenamefont {Ciampini}, \citenamefont {Morsch},\ and\
  \citenamefont {Arimondo}}]{eckardt2009exploring}%
  \BibitemOpen
  \bibfield  {author} {\bibinfo {author} {\bibfnamefont {A.}~\bibnamefont
  {Eckardt}}, \bibinfo {author} {\bibfnamefont {M.}~\bibnamefont {Holthaus}},
  \bibinfo {author} {\bibfnamefont {H.}~\bibnamefont {Lignier}}, \bibinfo
  {author} {\bibfnamefont {A.}~\bibnamefont {Zenesini}}, \bibinfo {author}
  {\bibfnamefont {D.}~\bibnamefont {Ciampini}}, \bibinfo {author}
  {\bibfnamefont {O.}~\bibnamefont {Morsch}},\ and\ \bibinfo {author}
  {\bibfnamefont {E.}~\bibnamefont {Arimondo}},\ }\bibfield  {title} {\bibinfo
  {title} {Exploring dynamic localization with a bose-einstein condensate},\
  }\href {https://doi.org/10.1103/PhysRevA.79.013611} {\bibfield  {journal}
  {\bibinfo  {journal} {Phys. Rev. A}\ }\textbf {\bibinfo {volume} {79}},\
  \bibinfo {pages} {013611} (\bibinfo {year} {2009})}\BibitemShut {NoStop}%
\bibitem [{\citenamefont {Bhakuni}\ and\ \citenamefont
  {Sharma}(2018)}]{bhakuni2018characteristic}%
  \BibitemOpen
  \bibfield  {author} {\bibinfo {author} {\bibfnamefont {D.~S.}\ \bibnamefont
  {Bhakuni}}\ and\ \bibinfo {author} {\bibfnamefont {A.}~\bibnamefont
  {Sharma}},\ }\bibfield  {title} {\bibinfo {title} {Characteristic length
  scales from entanglement dynamics in electric-field-driven tight-binding
  chains},\ }\href {https://doi.org/10.1103/PhysRevB.98.045408} {\bibfield
  {journal} {\bibinfo  {journal} {Phys. Rev. B}\ }\textbf {\bibinfo {volume}
  {98}},\ \bibinfo {pages} {045408} (\bibinfo {year} {2018})}\BibitemShut
  {NoStop}%
\bibitem [{\citenamefont {Holthaus}\ \emph
  {et~al.}(1995{\natexlab{a}})\citenamefont {Holthaus}, \citenamefont
  {Ristow},\ and\ \citenamefont {Hone}}]{holthaus1995random}%
  \BibitemOpen
  \bibfield  {author} {\bibinfo {author} {\bibfnamefont {M.}~\bibnamefont
  {Holthaus}}, \bibinfo {author} {\bibfnamefont {G.~H.}\ \bibnamefont
  {Ristow}},\ and\ \bibinfo {author} {\bibfnamefont {D.~W.}\ \bibnamefont
  {Hone}},\ }\bibfield  {title} {\bibinfo {title} {{Random Lattices in Combined
  a.c. and d.c. Electric Fields: Anderson vs. Wannier-Stark Localization}},\
  }\href {https://doi.org/10.1209/0295-5075/32/3/009} {\bibfield  {journal}
  {\bibinfo  {journal} {Europhysics Letters ({EPL})}\ }\textbf {\bibinfo
  {volume} {32}},\ \bibinfo {pages} {241} (\bibinfo {year}
  {1995}{\natexlab{a}})}\BibitemShut {NoStop}%
\bibitem [{\citenamefont {Holthaus}\ \emph
  {et~al.}(1995{\natexlab{b}})\citenamefont {Holthaus}, \citenamefont
  {Ristow},\ and\ \citenamefont {Hone}}]{holthaus1995ac}%
  \BibitemOpen
  \bibfield  {author} {\bibinfo {author} {\bibfnamefont {M.}~\bibnamefont
  {Holthaus}}, \bibinfo {author} {\bibfnamefont {G.~H.}\ \bibnamefont
  {Ristow}},\ and\ \bibinfo {author} {\bibfnamefont {D.~W.}\ \bibnamefont
  {Hone}},\ }\bibfield  {title} {\bibinfo {title} {ac-field-controlled anderson
  localization in disordered semiconductor superlattices},\ }\href
  {https://doi.org/10.1103/PhysRevLett.75.3914} {\bibfield  {journal} {\bibinfo
   {journal} {Phys. Rev. Lett.}\ }\textbf {\bibinfo {volume} {75}},\ \bibinfo
  {pages} {3914} (\bibinfo {year} {1995}{\natexlab{b}})}\BibitemShut {NoStop}%
\bibitem [{\citenamefont {Tiwari}\ \emph {et~al.}(2022)\citenamefont {Tiwari},
  \citenamefont {Bhakuni},\ and\ \citenamefont {Sharma}}]{tiwari2022noise}%
  \BibitemOpen
  \bibfield  {author} {\bibinfo {author} {\bibfnamefont {V.}~\bibnamefont
  {Tiwari}}, \bibinfo {author} {\bibfnamefont {D.~S.}\ \bibnamefont
  {Bhakuni}},\ and\ \bibinfo {author} {\bibfnamefont {A.}~\bibnamefont
  {Sharma}},\ }\bibfield  {title} {\bibinfo {title} {Noise-induced dynamical
  localization and delocalization},\ }\href
  {https://doi.org/10.1103/PhysRevB.105.165114} {\bibfield  {journal} {\bibinfo
   {journal} {Phys. Rev. B}\ }\textbf {\bibinfo {volume} {105}},\ \bibinfo
  {pages} {165114} (\bibinfo {year} {2022})}\BibitemShut {NoStop}%
\bibitem [{\citenamefont {Bhakuni}\ \emph {et~al.}(2020)\citenamefont
  {Bhakuni}, \citenamefont {Nehra},\ and\ \citenamefont
  {Sharma}}]{bhakuni2020drive}%
  \BibitemOpen
  \bibfield  {author} {\bibinfo {author} {\bibfnamefont {D.~S.}\ \bibnamefont
  {Bhakuni}}, \bibinfo {author} {\bibfnamefont {R.}~\bibnamefont {Nehra}},\
  and\ \bibinfo {author} {\bibfnamefont {A.}~\bibnamefont {Sharma}},\
  }\bibfield  {title} {\bibinfo {title} {{Drive-induced many-body localization
  and coherent destruction of Stark many-body localization}},\ }\href
  {https://doi.org/10.1103/PhysRevB.102.024201} {\bibfield  {journal} {\bibinfo
   {journal} {Phys. Rev. B}\ }\textbf {\bibinfo {volume} {102}},\ \bibinfo
  {pages} {024201} (\bibinfo {year} {2020})}\BibitemShut {NoStop}%
\bibitem [{\citenamefont {Kudo}\ and\ \citenamefont
  {Monteiro}(2011)}]{kudo2011theoretical}%
  \BibitemOpen
  \bibfield  {author} {\bibinfo {author} {\bibfnamefont {K.}~\bibnamefont
  {Kudo}}\ and\ \bibinfo {author} {\bibfnamefont {T.~S.}\ \bibnamefont
  {Monteiro}},\ }\bibfield  {title} {\bibinfo {title} {{Theoretical analysis of
  super-Bloch oscillations}},\ }\href
  {https://doi.org/10.1103/PhysRevA.83.053627} {\bibfield  {journal} {\bibinfo
  {journal} {Phys. Rev. A}\ }\textbf {\bibinfo {volume} {83}},\ \bibinfo
  {pages} {053627} (\bibinfo {year} {2011})}\BibitemShut {NoStop}%
\bibitem [{\citenamefont {Longhi}\ and\ \citenamefont
  {Della~Valle}(2012)}]{longhi2012correlated}%
  \BibitemOpen
  \bibfield  {author} {\bibinfo {author} {\bibfnamefont {S.}~\bibnamefont
  {Longhi}}\ and\ \bibinfo {author} {\bibfnamefont {G.}~\bibnamefont
  {Della~Valle}},\ }\bibfield  {title} {\bibinfo {title} {{Correlated
  super-Bloch oscillations}},\ }\href
  {https://doi.org/10.1103/PhysRevB.86.075143} {\bibfield  {journal} {\bibinfo
  {journal} {Phys. Rev. B}\ }\textbf {\bibinfo {volume} {86}},\ \bibinfo
  {pages} {075143} (\bibinfo {year} {2012})}\BibitemShut {NoStop}%
\bibitem [{\citenamefont {Schulz}\ \emph {et~al.}(2019)\citenamefont {Schulz},
  \citenamefont {Hooley}, \citenamefont {Moessner},\ and\ \citenamefont
  {Pollmann}}]{Schulz2019Stark}%
  \BibitemOpen
  \bibfield  {author} {\bibinfo {author} {\bibfnamefont {M.}~\bibnamefont
  {Schulz}}, \bibinfo {author} {\bibfnamefont {C.~A.}\ \bibnamefont {Hooley}},
  \bibinfo {author} {\bibfnamefont {R.}~\bibnamefont {Moessner}},\ and\
  \bibinfo {author} {\bibfnamefont {F.}~\bibnamefont {Pollmann}},\ }\bibfield
  {title} {\bibinfo {title} {{Stark Many-Body Localization}},\ }\href
  {https://doi.org/10.1103/PhysRevLett.122.040606} {\bibfield  {journal}
  {\bibinfo  {journal} {Phys. Rev. Lett.}\ }\textbf {\bibinfo {volume} {122}},\
  \bibinfo {pages} {040606} (\bibinfo {year} {2019})}\BibitemShut {NoStop}%
\bibitem [{\citenamefont {Bhakuni}\ and\ \citenamefont
  {Sharma}(2020{\natexlab{a}})}]{bhakuni2020entanglement}%
  \BibitemOpen
  \bibfield  {author} {\bibinfo {author} {\bibfnamefont {D.~S.}\ \bibnamefont
  {Bhakuni}}\ and\ \bibinfo {author} {\bibfnamefont {A.}~\bibnamefont
  {Sharma}},\ }\bibfield  {title} {\bibinfo {title} {Entanglement and
  thermodynamic entropy in a clean many-body-localized system},\ }\href
  {https://doi.org/10.1088/1361-648X/ab7c92} {\bibfield  {journal} {\bibinfo
  {journal} {Journal of Physics: Condensed Matter}\ }\textbf {\bibinfo {volume}
  {32}},\ \bibinfo {pages} {255603} (\bibinfo {year}
  {2020}{\natexlab{a}})}\BibitemShut {NoStop}%
\bibitem [{\citenamefont {Morong}\ \emph {et~al.}(2021)\citenamefont {Morong},
  \citenamefont {Liu}, \citenamefont {Becker}, \citenamefont {Collins},
  \citenamefont {Feng}, \citenamefont {Kyprianidis}, \citenamefont {Pagano},
  \citenamefont {You}, \citenamefont {Gorshkov},\ and\ \citenamefont
  {Monroe}}]{morong2021observation}%
  \BibitemOpen
  \bibfield  {author} {\bibinfo {author} {\bibfnamefont {W.}~\bibnamefont
  {Morong}}, \bibinfo {author} {\bibfnamefont {F.}~\bibnamefont {Liu}},
  \bibinfo {author} {\bibfnamefont {P.}~\bibnamefont {Becker}}, \bibinfo
  {author} {\bibfnamefont {K.~S.}\ \bibnamefont {Collins}}, \bibinfo {author}
  {\bibfnamefont {L.}~\bibnamefont {Feng}}, \bibinfo {author} {\bibfnamefont
  {A.}~\bibnamefont {Kyprianidis}}, \bibinfo {author} {\bibfnamefont
  {G.}~\bibnamefont {Pagano}}, \bibinfo {author} {\bibfnamefont
  {T.}~\bibnamefont {You}}, \bibinfo {author} {\bibfnamefont {A.~V.}\
  \bibnamefont {Gorshkov}},\ and\ \bibinfo {author} {\bibfnamefont
  {C.}~\bibnamefont {Monroe}},\ }\bibfield  {title} {\bibinfo {title}
  {Observation of stark many-body localization without disorder},\ }\href
  {https://doi.org/10.1038/s41586-021-03988-0} {\bibfield  {journal} {\bibinfo
  {journal} {Nature}\ }\textbf {\bibinfo {volume} {599}},\ \bibinfo {pages}
  {393} (\bibinfo {year} {2021})}\BibitemShut {NoStop}%
\bibitem [{\citenamefont {Guo}\ \emph {et~al.}(2021)\citenamefont {Guo},
  \citenamefont {Cheng}, \citenamefont {Li}, \citenamefont {Xu}, \citenamefont
  {Zhang}, \citenamefont {Wang}, \citenamefont {Song}, \citenamefont {Liu},
  \citenamefont {Ren}, \citenamefont {Dong}, \citenamefont {Mondaini},\ and\
  \citenamefont {Wang}}]{guo2020stark}%
  \BibitemOpen
  \bibfield  {author} {\bibinfo {author} {\bibfnamefont {Q.}~\bibnamefont
  {Guo}}, \bibinfo {author} {\bibfnamefont {C.}~\bibnamefont {Cheng}}, \bibinfo
  {author} {\bibfnamefont {H.}~\bibnamefont {Li}}, \bibinfo {author}
  {\bibfnamefont {S.}~\bibnamefont {Xu}}, \bibinfo {author} {\bibfnamefont
  {P.}~\bibnamefont {Zhang}}, \bibinfo {author} {\bibfnamefont
  {Z.}~\bibnamefont {Wang}}, \bibinfo {author} {\bibfnamefont {C.}~\bibnamefont
  {Song}}, \bibinfo {author} {\bibfnamefont {W.}~\bibnamefont {Liu}}, \bibinfo
  {author} {\bibfnamefont {W.}~\bibnamefont {Ren}}, \bibinfo {author}
  {\bibfnamefont {H.}~\bibnamefont {Dong}}, \bibinfo {author} {\bibfnamefont
  {R.}~\bibnamefont {Mondaini}},\ and\ \bibinfo {author} {\bibfnamefont
  {H.}~\bibnamefont {Wang}},\ }\bibfield  {title} {\bibinfo {title} {Stark
  many-body localization on a superconducting quantum processor},\ }\href
  {https://doi.org/10.1103/PhysRevLett.127.240502} {\bibfield  {journal}
  {\bibinfo  {journal} {Phys. Rev. Lett.}\ }\textbf {\bibinfo {volume} {127}},\
  \bibinfo {pages} {240502} (\bibinfo {year} {2021})}\BibitemShut {NoStop}%
\bibitem [{\citenamefont {Taylor}\ \emph {et~al.}(2020)\citenamefont {Taylor},
  \citenamefont {Schulz}, \citenamefont {Pollmann},\ and\ \citenamefont
  {Moessner}}]{taylor2020experimental}%
  \BibitemOpen
  \bibfield  {author} {\bibinfo {author} {\bibfnamefont {S.~R.}\ \bibnamefont
  {Taylor}}, \bibinfo {author} {\bibfnamefont {M.}~\bibnamefont {Schulz}},
  \bibinfo {author} {\bibfnamefont {F.}~\bibnamefont {Pollmann}},\ and\
  \bibinfo {author} {\bibfnamefont {R.}~\bibnamefont {Moessner}},\ }\bibfield
  {title} {\bibinfo {title} {Experimental probes of stark many-body
  localization},\ }\href {https://doi.org/10.1103/PhysRevB.102.054206}
  {\bibfield  {journal} {\bibinfo  {journal} {Phys. Rev. B}\ }\textbf {\bibinfo
  {volume} {102}},\ \bibinfo {pages} {054206} (\bibinfo {year}
  {2020})}\BibitemShut {NoStop}%
\bibitem [{\citenamefont {van Nieuwenburg}\ \emph {et~al.}(2019)\citenamefont
  {van Nieuwenburg}, \citenamefont {Baum},\ and\ \citenamefont
  {Refael}}]{evert2019bloch}%
  \BibitemOpen
  \bibfield  {author} {\bibinfo {author} {\bibfnamefont {E.}~\bibnamefont {van
  Nieuwenburg}}, \bibinfo {author} {\bibfnamefont {Y.}~\bibnamefont {Baum}},\
  and\ \bibinfo {author} {\bibfnamefont {G.}~\bibnamefont {Refael}},\
  }\bibfield  {title} {\bibinfo {title} {{From Bloch oscillations to many-body
  localization in clean interacting systems}},\ }\href
  {https://doi.org/10.1073/pnas.1819316116} {\bibfield  {journal} {\bibinfo
  {journal} {Proceedings of the National Academy of Sciences}\ }\textbf
  {\bibinfo {volume} {116}},\ \bibinfo {pages} {9269} (\bibinfo {year}
  {2019})}\BibitemShut {NoStop}%
\bibitem [{\citenamefont {Bhakuni}\ and\ \citenamefont
  {Sharma}(2020{\natexlab{b}})}]{bhakuni2020stability}%
  \BibitemOpen
  \bibfield  {author} {\bibinfo {author} {\bibfnamefont {D.~S.}\ \bibnamefont
  {Bhakuni}}\ and\ \bibinfo {author} {\bibfnamefont {A.}~\bibnamefont
  {Sharma}},\ }\bibfield  {title} {\bibinfo {title} {Stability of electric
  field driven many-body localization in an interacting long-range hopping
  model},\ }\href {https://doi.org/10.1103/PhysRevB.102.085133} {\bibfield
  {journal} {\bibinfo  {journal} {Phys. Rev. B}\ }\textbf {\bibinfo {volume}
  {102}},\ \bibinfo {pages} {085133} (\bibinfo {year}
  {2020}{\natexlab{b}})}\BibitemShut {NoStop}%
\bibitem [{\citenamefont {Ribeiro}\ \emph {et~al.}(2020)\citenamefont
  {Ribeiro}, \citenamefont {Lazarides},\ and\ \citenamefont
  {Haque}}]{ribeiro2020many}%
  \BibitemOpen
  \bibfield  {author} {\bibinfo {author} {\bibfnamefont {P.}~\bibnamefont
  {Ribeiro}}, \bibinfo {author} {\bibfnamefont {A.}~\bibnamefont {Lazarides}},\
  and\ \bibinfo {author} {\bibfnamefont {M.}~\bibnamefont {Haque}},\ }\bibfield
   {title} {\bibinfo {title} {{Many-Body Quantum Dynamics of Initially Trapped
  Systems due to a Stark Potential: Thermalization versus Bloch
  Oscillations}},\ }\href {https://doi.org/10.1103/PhysRevLett.124.110603}
  {\bibfield  {journal} {\bibinfo  {journal} {Phys. Rev. Lett.}\ }\textbf
  {\bibinfo {volume} {124}},\ \bibinfo {pages} {110603} (\bibinfo {year}
  {2020})}\BibitemShut {NoStop}%
\bibitem [{\citenamefont {Doggen}\ \emph {et~al.}(2021)\citenamefont {Doggen},
  \citenamefont {Gornyi},\ and\ \citenamefont {Polyakov}}]{doggen2021stark}%
  \BibitemOpen
  \bibfield  {author} {\bibinfo {author} {\bibfnamefont {E.~V.~H.}\
  \bibnamefont {Doggen}}, \bibinfo {author} {\bibfnamefont {I.~V.}\
  \bibnamefont {Gornyi}},\ and\ \bibinfo {author} {\bibfnamefont {D.~G.}\
  \bibnamefont {Polyakov}},\ }\bibfield  {title} {\bibinfo {title} {Stark
  many-body localization: Evidence for hilbert-space shattering},\ }\href
  {https://doi.org/10.1103/PhysRevB.103.L100202} {\bibfield  {journal}
  {\bibinfo  {journal} {Phys. Rev. B}\ }\textbf {\bibinfo {volume} {103}},\
  \bibinfo {pages} {L100202} (\bibinfo {year} {2021})}\BibitemShut {NoStop}%
\bibitem [{\citenamefont {Zisling}\ \emph {et~al.}(2022)\citenamefont
  {Zisling}, \citenamefont {Kennes},\ and\ \citenamefont
  {Bar~Lev}}]{zisling2022transport}%
  \BibitemOpen
  \bibfield  {author} {\bibinfo {author} {\bibfnamefont {G.}~\bibnamefont
  {Zisling}}, \bibinfo {author} {\bibfnamefont {D.~M.}\ \bibnamefont
  {Kennes}},\ and\ \bibinfo {author} {\bibfnamefont {Y.}~\bibnamefont
  {Bar~Lev}},\ }\bibfield  {title} {\bibinfo {title} {Transport in stark
  many-body localized systems},\ }\href
  {https://doi.org/10.1103/PhysRevB.105.L140201} {\bibfield  {journal}
  {\bibinfo  {journal} {Phys. Rev. B}\ }\textbf {\bibinfo {volume} {105}},\
  \bibinfo {pages} {L140201} (\bibinfo {year} {2022})}\BibitemShut {NoStop}%
\bibitem [{\citenamefont {Bairey}\ \emph {et~al.}(2017)\citenamefont {Bairey},
  \citenamefont {Refael},\ and\ \citenamefont {Lindner}}]{bairey2017driving}%
  \BibitemOpen
  \bibfield  {author} {\bibinfo {author} {\bibfnamefont {E.}~\bibnamefont
  {Bairey}}, \bibinfo {author} {\bibfnamefont {G.}~\bibnamefont {Refael}},\
  and\ \bibinfo {author} {\bibfnamefont {N.~H.}\ \bibnamefont {Lindner}},\
  }\bibfield  {title} {\bibinfo {title} {Driving induced many-body
  localization},\ }\href {https://doi.org/10.1103/PhysRevB.96.020201}
  {\bibfield  {journal} {\bibinfo  {journal} {Phys. Rev. B}\ }\textbf {\bibinfo
  {volume} {96}},\ \bibinfo {pages} {020201} (\bibinfo {year}
  {2017})}\BibitemShut {NoStop}%
\bibitem [{\citenamefont {Kshetrimayum}\ \emph {et~al.}(2020)\citenamefont
  {Kshetrimayum}, \citenamefont {Eisert},\ and\ \citenamefont
  {Kennes}}]{kshetrimayum2020stark}%
  \BibitemOpen
  \bibfield  {author} {\bibinfo {author} {\bibfnamefont {A.}~\bibnamefont
  {Kshetrimayum}}, \bibinfo {author} {\bibfnamefont {J.}~\bibnamefont
  {Eisert}},\ and\ \bibinfo {author} {\bibfnamefont {D.~M.}\ \bibnamefont
  {Kennes}},\ }\bibfield  {title} {\bibinfo {title} {Stark time crystals:
  Symmetry breaking in space and time},\ }\href
  {https://doi.org/10.1103/PhysRevB.102.195116} {\bibfield  {journal} {\bibinfo
   {journal} {Phys. Rev. B}\ }\textbf {\bibinfo {volume} {102}},\ \bibinfo
  {pages} {195116} (\bibinfo {year} {2020})}\BibitemShut {NoStop}%
\bibitem [{\citenamefont {Liu}\ \emph {et~al.}(2023)\citenamefont {Liu},
  \citenamefont {Zhang}, \citenamefont {Hsieh}, \citenamefont {Zhang},\ and\
  \citenamefont {Yao}}]{liu2022discrete}%
  \BibitemOpen
  \bibfield  {author} {\bibinfo {author} {\bibfnamefont {S.}~\bibnamefont
  {Liu}}, \bibinfo {author} {\bibfnamefont {S.-X.}\ \bibnamefont {Zhang}},
  \bibinfo {author} {\bibfnamefont {C.-Y.}\ \bibnamefont {Hsieh}}, \bibinfo
  {author} {\bibfnamefont {S.}~\bibnamefont {Zhang}},\ and\ \bibinfo {author}
  {\bibfnamefont {H.}~\bibnamefont {Yao}},\ }\bibfield  {title} {\bibinfo
  {title} {{Discrete Time Crystal Enabled by Stark Many-Body Localization}},\
  }\href {https://doi.org/10.1103/PhysRevLett.130.120403} {\bibfield  {journal}
  {\bibinfo  {journal} {Phys. Rev. Lett.}\ }\textbf {\bibinfo {volume} {130}},\
  \bibinfo {pages} {120403} (\bibinfo {year} {2023})}\BibitemShut {NoStop}%
\bibitem [{\citenamefont {Abal}\ \emph {et~al.}(2002)\citenamefont {Abal},
  \citenamefont {Donangelo}, \citenamefont {Romanelli}, \citenamefont
  {Sicardi~Schifino},\ and\ \citenamefont {Siri}}]{abal2002dynamical}%
  \BibitemOpen
  \bibfield  {author} {\bibinfo {author} {\bibfnamefont {G.}~\bibnamefont
  {Abal}}, \bibinfo {author} {\bibfnamefont {R.}~\bibnamefont {Donangelo}},
  \bibinfo {author} {\bibfnamefont {A.}~\bibnamefont {Romanelli}}, \bibinfo
  {author} {\bibfnamefont {A.~C.}\ \bibnamefont {Sicardi~Schifino}},\ and\
  \bibinfo {author} {\bibfnamefont {R.}~\bibnamefont {Siri}},\ }\bibfield
  {title} {\bibinfo {title} {Dynamical localization in quasiperiodic driven
  systems},\ }\href {https://doi.org/10.1103/PhysRevE.65.046236} {\bibfield
  {journal} {\bibinfo  {journal} {Phys. Rev. E}\ }\textbf {\bibinfo {volume}
  {65}},\ \bibinfo {pages} {046236} (\bibinfo {year} {2002})}\BibitemShut
  {NoStop}%
\bibitem [{\citenamefont {Verdeny}\ \emph {et~al.}(2016)\citenamefont
  {Verdeny}, \citenamefont {Puig},\ and\ \citenamefont
  {Mintert}}]{verdeny2016quasi}%
  \BibitemOpen
  \bibfield  {author} {\bibinfo {author} {\bibfnamefont {A.}~\bibnamefont
  {Verdeny}}, \bibinfo {author} {\bibfnamefont {J.}~\bibnamefont {Puig}},\ and\
  \bibinfo {author} {\bibfnamefont {F.}~\bibnamefont {Mintert}},\ }\bibfield
  {title} {\bibinfo {title} {Quasi-periodically driven quantum systems},\
  }\href {https://doi.org/doi:10.1515/zna-2016-0079} {\bibfield  {journal}
  {\bibinfo  {journal} {Zeitschrift für Naturforschung A}\ }\textbf {\bibinfo
  {volume} {71}},\ \bibinfo {pages} {897} (\bibinfo {year} {2016})}\BibitemShut
  {NoStop}%
\bibitem [{\citenamefont {Nandy}\ \emph {et~al.}(2017)\citenamefont {Nandy},
  \citenamefont {Sen},\ and\ \citenamefont {Sen}}]{nandy2017aperiodically}%
  \BibitemOpen
  \bibfield  {author} {\bibinfo {author} {\bibfnamefont {S.}~\bibnamefont
  {Nandy}}, \bibinfo {author} {\bibfnamefont {A.}~\bibnamefont {Sen}},\ and\
  \bibinfo {author} {\bibfnamefont {D.}~\bibnamefont {Sen}},\ }\bibfield
  {title} {\bibinfo {title} {Aperiodically driven integrable systems and their
  emergent steady states},\ }\href {https://doi.org/10.1103/PhysRevX.7.031034}
  {\bibfield  {journal} {\bibinfo  {journal} {Phys. Rev. X}\ }\textbf {\bibinfo
  {volume} {7}},\ \bibinfo {pages} {031034} (\bibinfo {year}
  {2017})}\BibitemShut {NoStop}%
\bibitem [{\citenamefont {Dumitrescu}\ \emph {et~al.}(2018)\citenamefont
  {Dumitrescu}, \citenamefont {Vasseur},\ and\ \citenamefont
  {Potter}}]{dumitrescu2018logarithmic}%
  \BibitemOpen
  \bibfield  {author} {\bibinfo {author} {\bibfnamefont {P.~T.}\ \bibnamefont
  {Dumitrescu}}, \bibinfo {author} {\bibfnamefont {R.}~\bibnamefont
  {Vasseur}},\ and\ \bibinfo {author} {\bibfnamefont {A.~C.}\ \bibnamefont
  {Potter}},\ }\bibfield  {title} {\bibinfo {title} {Logarithmically slow
  relaxation in quasiperiodically driven random spin chains},\ }\href
  {https://doi.org/10.1103/PhysRevLett.120.070602} {\bibfield  {journal}
  {\bibinfo  {journal} {Phys. Rev. Lett.}\ }\textbf {\bibinfo {volume} {120}},\
  \bibinfo {pages} {070602} (\bibinfo {year} {2018})}\BibitemShut {NoStop}%
\bibitem [{\citenamefont {Nandy}\ \emph {et~al.}(2018)\citenamefont {Nandy},
  \citenamefont {Sen},\ and\ \citenamefont {Sen}}]{nandy2018steady}%
  \BibitemOpen
  \bibfield  {author} {\bibinfo {author} {\bibfnamefont {S.}~\bibnamefont
  {Nandy}}, \bibinfo {author} {\bibfnamefont {A.}~\bibnamefont {Sen}},\ and\
  \bibinfo {author} {\bibfnamefont {D.}~\bibnamefont {Sen}},\ }\bibfield
  {title} {\bibinfo {title} {Steady states of a quasiperiodically driven
  integrable system},\ }\href {https://doi.org/10.1103/PhysRevB.98.245144}
  {\bibfield  {journal} {\bibinfo  {journal} {Phys. Rev. B}\ }\textbf {\bibinfo
  {volume} {98}},\ \bibinfo {pages} {245144} (\bibinfo {year}
  {2018})}\BibitemShut {NoStop}%
\bibitem [{\citenamefont {Giergiel}\ \emph {et~al.}(2019)\citenamefont
  {Giergiel}, \citenamefont {Kuro\ifmmode~\acute{s}\else \'{s}\fi{}},\ and\
  \citenamefont {Sacha}}]{giergiel2019discrete}%
  \BibitemOpen
  \bibfield  {author} {\bibinfo {author} {\bibfnamefont {K.}~\bibnamefont
  {Giergiel}}, \bibinfo {author} {\bibfnamefont {A.}~\bibnamefont
  {Kuro\ifmmode~\acute{s}\else \'{s}\fi{}}},\ and\ \bibinfo {author}
  {\bibfnamefont {K.}~\bibnamefont {Sacha}},\ }\bibfield  {title} {\bibinfo
  {title} {Discrete time quasicrystals},\ }\href
  {https://doi.org/10.1103/PhysRevB.99.220303} {\bibfield  {journal} {\bibinfo
  {journal} {Phys. Rev. B}\ }\textbf {\bibinfo {volume} {99}},\ \bibinfo
  {pages} {220303} (\bibinfo {year} {2019})}\BibitemShut {NoStop}%
\bibitem [{\citenamefont {Ray}\ \emph {et~al.}(2019)\citenamefont {Ray},
  \citenamefont {Sinha},\ and\ \citenamefont {Sen}}]{ray2019dynamics}%
  \BibitemOpen
  \bibfield  {author} {\bibinfo {author} {\bibfnamefont {S.}~\bibnamefont
  {Ray}}, \bibinfo {author} {\bibfnamefont {S.}~\bibnamefont {Sinha}},\ and\
  \bibinfo {author} {\bibfnamefont {D.}~\bibnamefont {Sen}},\ }\bibfield
  {title} {\bibinfo {title} {Dynamics of quasiperiodically driven spin
  systems},\ }\href {https://doi.org/10.1103/PhysRevE.100.052129} {\bibfield
  {journal} {\bibinfo  {journal} {Phys. Rev. E}\ }\textbf {\bibinfo {volume}
  {100}},\ \bibinfo {pages} {052129} (\bibinfo {year} {2019})}\BibitemShut
  {NoStop}%
\bibitem [{\citenamefont {Maity}\ \emph {et~al.}(2019)\citenamefont {Maity},
  \citenamefont {Bhattacharya}, \citenamefont {Dutta},\ and\ \citenamefont
  {Sen}}]{maity2019fibonacci}%
  \BibitemOpen
  \bibfield  {author} {\bibinfo {author} {\bibfnamefont {S.}~\bibnamefont
  {Maity}}, \bibinfo {author} {\bibfnamefont {U.}~\bibnamefont {Bhattacharya}},
  \bibinfo {author} {\bibfnamefont {A.}~\bibnamefont {Dutta}},\ and\ \bibinfo
  {author} {\bibfnamefont {D.}~\bibnamefont {Sen}},\ }\bibfield  {title}
  {\bibinfo {title} {Fibonacci steady states in a driven integrable quantum
  system},\ }\href {https://doi.org/10.1103/PhysRevB.99.020306} {\bibfield
  {journal} {\bibinfo  {journal} {Phys. Rev. B}\ }\textbf {\bibinfo {volume}
  {99}},\ \bibinfo {pages} {020306} (\bibinfo {year} {2019})}\BibitemShut
  {NoStop}%
\bibitem [{\citenamefont {Else}\ \emph {et~al.}(2020)\citenamefont {Else},
  \citenamefont {Ho},\ and\ \citenamefont {Dumitrescu}}]{else2020long}%
  \BibitemOpen
  \bibfield  {author} {\bibinfo {author} {\bibfnamefont {D.~V.}\ \bibnamefont
  {Else}}, \bibinfo {author} {\bibfnamefont {W.~W.}\ \bibnamefont {Ho}},\ and\
  \bibinfo {author} {\bibfnamefont {P.~T.}\ \bibnamefont {Dumitrescu}},\
  }\bibfield  {title} {\bibinfo {title} {Long-lived interacting phases of
  matter protected by multiple time-translation symmetries in quasiperiodically
  driven systems},\ }\href {https://doi.org/10.1103/PhysRevX.10.021032}
  {\bibfield  {journal} {\bibinfo  {journal} {Phys. Rev. X}\ }\textbf {\bibinfo
  {volume} {10}},\ \bibinfo {pages} {021032} (\bibinfo {year}
  {2020})}\BibitemShut {NoStop}%
\bibitem [{\citenamefont {Zhao}\ \emph {et~al.}(2021)\citenamefont {Zhao},
  \citenamefont {Mintert}, \citenamefont {Moessner},\ and\ \citenamefont
  {Knolle}}]{zhao2021random}%
  \BibitemOpen
  \bibfield  {author} {\bibinfo {author} {\bibfnamefont {H.}~\bibnamefont
  {Zhao}}, \bibinfo {author} {\bibfnamefont {F.}~\bibnamefont {Mintert}},
  \bibinfo {author} {\bibfnamefont {R.}~\bibnamefont {Moessner}},\ and\
  \bibinfo {author} {\bibfnamefont {J.}~\bibnamefont {Knolle}},\ }\bibfield
  {title} {\bibinfo {title} {Random multipolar driving: Tunably slow heating
  through spectral engineering},\ }\href
  {https://doi.org/10.1103/PhysRevLett.126.040601} {\bibfield  {journal}
  {\bibinfo  {journal} {Phys. Rev. Lett.}\ }\textbf {\bibinfo {volume} {126}},\
  \bibinfo {pages} {040601} (\bibinfo {year} {2021})}\BibitemShut {NoStop}%
\bibitem [{\citenamefont {Mori}\ \emph {et~al.}(2021)\citenamefont {Mori},
  \citenamefont {Zhao}, \citenamefont {Mintert}, \citenamefont {Knolle},\ and\
  \citenamefont {Moessner}}]{mori2021rigorous}%
  \BibitemOpen
  \bibfield  {author} {\bibinfo {author} {\bibfnamefont {T.}~\bibnamefont
  {Mori}}, \bibinfo {author} {\bibfnamefont {H.}~\bibnamefont {Zhao}}, \bibinfo
  {author} {\bibfnamefont {F.}~\bibnamefont {Mintert}}, \bibinfo {author}
  {\bibfnamefont {J.}~\bibnamefont {Knolle}},\ and\ \bibinfo {author}
  {\bibfnamefont {R.}~\bibnamefont {Moessner}},\ }\bibfield  {title} {\bibinfo
  {title} {Rigorous bounds on the heating rate in thue-morse quasiperiodically
  and randomly driven quantum many-body systems},\ }\href
  {https://doi.org/10.1103/PhysRevLett.127.050602} {\bibfield  {journal}
  {\bibinfo  {journal} {Phys. Rev. Lett.}\ }\textbf {\bibinfo {volume} {127}},\
  \bibinfo {pages} {050602} (\bibinfo {year} {2021})}\BibitemShut {NoStop}%
\bibitem [{\citenamefont {Zhao}\ \emph
  {et~al.}(2022{\natexlab{a}})\citenamefont {Zhao}, \citenamefont {Mintert},
  \citenamefont {Knolle},\ and\ \citenamefont
  {Moessner}}]{zhao2022localization}%
  \BibitemOpen
  \bibfield  {author} {\bibinfo {author} {\bibfnamefont {H.}~\bibnamefont
  {Zhao}}, \bibinfo {author} {\bibfnamefont {F.}~\bibnamefont {Mintert}},
  \bibinfo {author} {\bibfnamefont {J.}~\bibnamefont {Knolle}},\ and\ \bibinfo
  {author} {\bibfnamefont {R.}~\bibnamefont {Moessner}},\ }\bibfield  {title}
  {\bibinfo {title} {Localization persisting under aperiodic driving},\ }\href
  {https://doi.org/10.1103/PhysRevB.105.L220202} {\bibfield  {journal}
  {\bibinfo  {journal} {Phys. Rev. B}\ }\textbf {\bibinfo {volume} {105}},\
  \bibinfo {pages} {L220202} (\bibinfo {year}
  {2022}{\natexlab{a}})}\BibitemShut {NoStop}%
\bibitem [{\citenamefont {Long}\ \emph {et~al.}(2022)\citenamefont {Long},
  \citenamefont {Crowley},\ and\ \citenamefont {Chandran}}]{long2022many}%
  \BibitemOpen
  \bibfield  {author} {\bibinfo {author} {\bibfnamefont {D.~M.}\ \bibnamefont
  {Long}}, \bibinfo {author} {\bibfnamefont {P.~J.~D.}\ \bibnamefont
  {Crowley}},\ and\ \bibinfo {author} {\bibfnamefont {A.}~\bibnamefont
  {Chandran}},\ }\bibfield  {title} {\bibinfo {title} {Many-body localization
  with quasiperiodic driving},\ }\href
  {https://doi.org/10.1103/PhysRevB.105.144204} {\bibfield  {journal} {\bibinfo
   {journal} {Phys. Rev. B}\ }\textbf {\bibinfo {volume} {105}},\ \bibinfo
  {pages} {144204} (\bibinfo {year} {2022})}\BibitemShut {NoStop}%
\bibitem [{\citenamefont {Martin}\ \emph {et~al.}(2022)\citenamefont {Martin},
  \citenamefont {Martin},\ and\ \citenamefont {Agarwal}}]{martin2022effect}%
  \BibitemOpen
  \bibfield  {author} {\bibinfo {author} {\bibfnamefont {T.}~\bibnamefont
  {Martin}}, \bibinfo {author} {\bibfnamefont {I.}~\bibnamefont {Martin}},\
  and\ \bibinfo {author} {\bibfnamefont {K.}~\bibnamefont {Agarwal}},\
  }\bibfield  {title} {\bibinfo {title} {Effect of quasiperiodic and random
  noise on many-body dynamical decoupling protocols},\ }\href
  {https://doi.org/10.1103/PhysRevB.106.134306} {\bibfield  {journal} {\bibinfo
   {journal} {Phys. Rev. B}\ }\textbf {\bibinfo {volume} {106}},\ \bibinfo
  {pages} {134306} (\bibinfo {year} {2022})}\BibitemShut {NoStop}%
\bibitem [{\citenamefont {Zhao}\ \emph
  {et~al.}(2022{\natexlab{b}})\citenamefont {Zhao}, \citenamefont {Knolle},
  \citenamefont {Moessner},\ and\ \citenamefont
  {Mintert}}]{zhao2022suppression}%
  \BibitemOpen
  \bibfield  {author} {\bibinfo {author} {\bibfnamefont {H.}~\bibnamefont
  {Zhao}}, \bibinfo {author} {\bibfnamefont {J.}~\bibnamefont {Knolle}},
  \bibinfo {author} {\bibfnamefont {R.}~\bibnamefont {Moessner}},\ and\
  \bibinfo {author} {\bibfnamefont {F.}~\bibnamefont {Mintert}},\ }\bibfield
  {title} {\bibinfo {title} {Suppression of interband heating for random
  driving},\ }\href {https://doi.org/10.1103/PhysRevLett.129.120605} {\bibfield
   {journal} {\bibinfo  {journal} {Phys. Rev. Lett.}\ }\textbf {\bibinfo
  {volume} {129}},\ \bibinfo {pages} {120605} (\bibinfo {year}
  {2022}{\natexlab{b}})}\BibitemShut {NoStop}%
\bibitem [{\citenamefont {Zhao}\ \emph
  {et~al.}(2022{\natexlab{c}})\citenamefont {Zhao}, \citenamefont {Rudner},
  \citenamefont {Moessner},\ and\ \citenamefont {Knolle}}]{zhao2022anomalous}%
  \BibitemOpen
  \bibfield  {author} {\bibinfo {author} {\bibfnamefont {H.}~\bibnamefont
  {Zhao}}, \bibinfo {author} {\bibfnamefont {M.~S.}\ \bibnamefont {Rudner}},
  \bibinfo {author} {\bibfnamefont {R.}~\bibnamefont {Moessner}},\ and\
  \bibinfo {author} {\bibfnamefont {J.}~\bibnamefont {Knolle}},\ }\bibfield
  {title} {\bibinfo {title} {Anomalous random multipolar driven insulators},\
  }\href {https://doi.org/10.1103/PhysRevB.105.245119} {\bibfield  {journal}
  {\bibinfo  {journal} {Phys. Rev. B}\ }\textbf {\bibinfo {volume} {105}},\
  \bibinfo {pages} {245119} (\bibinfo {year} {2022}{\natexlab{c}})}\BibitemShut
  {NoStop}%
\bibitem [{\citenamefont {Das}\ \emph {et~al.}(2023)\citenamefont {Das},
  \citenamefont {Singh~Bhakuni}, \citenamefont {Santos},\ and\ \citenamefont
  {Sharma}}]{das2023periodically}%
  \BibitemOpen
  \bibfield  {author} {\bibinfo {author} {\bibfnamefont {P.}~\bibnamefont
  {Das}}, \bibinfo {author} {\bibfnamefont {D.}~\bibnamefont {Singh~Bhakuni}},
  \bibinfo {author} {\bibfnamefont {L.~F.}\ \bibnamefont {Santos}},\ and\
  \bibinfo {author} {\bibfnamefont {A.}~\bibnamefont {Sharma}},\ }\bibfield
  {title} {\bibinfo {title} {Periodically and quasiperiodically
  driven-anisotropic dicke model},\ }\href@noop {} {\bibfield  {journal}
  {\bibinfo  {journal} {arXiv e-prints}\ ,\ \bibinfo {pages} {arXiv}} (\bibinfo
  {year} {2023})}\BibitemShut {NoStop}%
\bibitem [{\citenamefont {Mukherjee}\ \emph {et~al.}(2020)\citenamefont
  {Mukherjee}, \citenamefont {Sen}, \citenamefont {Sen},\ and\ \citenamefont
  {Sengupta}}]{mukherjee2020restoring}%
  \BibitemOpen
  \bibfield  {author} {\bibinfo {author} {\bibfnamefont {B.}~\bibnamefont
  {Mukherjee}}, \bibinfo {author} {\bibfnamefont {A.}~\bibnamefont {Sen}},
  \bibinfo {author} {\bibfnamefont {D.}~\bibnamefont {Sen}},\ and\ \bibinfo
  {author} {\bibfnamefont {K.}~\bibnamefont {Sengupta}},\ }\bibfield  {title}
  {\bibinfo {title} {Restoring coherence via aperiodic drives in a many-body
  quantum system},\ }\href {https://doi.org/10.1103/PhysRevB.102.014301}
  {\bibfield  {journal} {\bibinfo  {journal} {Phys. Rev. B}\ }\textbf {\bibinfo
  {volume} {102}},\ \bibinfo {pages} {014301} (\bibinfo {year}
  {2020})}\BibitemShut {NoStop}%
\bibitem [{\citenamefont {Zhao}\ \emph {et~al.}(2019)\citenamefont {Zhao},
  \citenamefont {Mintert},\ and\ \citenamefont {Knolle}}]{zhao2019floquet}%
  \BibitemOpen
  \bibfield  {author} {\bibinfo {author} {\bibfnamefont {H.}~\bibnamefont
  {Zhao}}, \bibinfo {author} {\bibfnamefont {F.}~\bibnamefont {Mintert}},\ and\
  \bibinfo {author} {\bibfnamefont {J.}~\bibnamefont {Knolle}},\ }\bibfield
  {title} {\bibinfo {title} {{Floquet time spirals and stable discrete-time
  quasicrystals in quasiperiodically driven quantum many-body systems}},\
  }\href {https://doi.org/10.1103/PhysRevB.100.134302} {\bibfield  {journal}
  {\bibinfo  {journal} {Phys. Rev. B}\ }\textbf {\bibinfo {volume} {100}},\
  \bibinfo {pages} {134302} (\bibinfo {year} {2019})}\BibitemShut {NoStop}%
\bibitem [{\citenamefont {Dumitrescu}\ \emph {et~al.}(2022)\citenamefont
  {Dumitrescu}, \citenamefont {Bohnet}, \citenamefont {Gaebler}, \citenamefont
  {Hankin}, \citenamefont {Hayes}, \citenamefont {Kumar}, \citenamefont
  {Neyenhuis}, \citenamefont {Vasseur},\ and\ \citenamefont
  {Potter}}]{dumitrescu2022dynamical}%
  \BibitemOpen
  \bibfield  {author} {\bibinfo {author} {\bibfnamefont {P.~T.}\ \bibnamefont
  {Dumitrescu}}, \bibinfo {author} {\bibfnamefont {J.~G.}\ \bibnamefont
  {Bohnet}}, \bibinfo {author} {\bibfnamefont {J.~P.}\ \bibnamefont {Gaebler}},
  \bibinfo {author} {\bibfnamefont {A.}~\bibnamefont {Hankin}}, \bibinfo
  {author} {\bibfnamefont {D.}~\bibnamefont {Hayes}}, \bibinfo {author}
  {\bibfnamefont {A.}~\bibnamefont {Kumar}}, \bibinfo {author} {\bibfnamefont
  {B.}~\bibnamefont {Neyenhuis}}, \bibinfo {author} {\bibfnamefont
  {R.}~\bibnamefont {Vasseur}},\ and\ \bibinfo {author} {\bibfnamefont {A.~C.}\
  \bibnamefont {Potter}},\ }\bibfield  {title} {\bibinfo {title} {Dynamical
  topological phase realized in a trapped-ion quantum simulator},\ }\href
  {https://doi.org/10.1038/s41586-022-04853-4} {\bibfield  {journal} {\bibinfo
  {journal} {Nature}\ }\textbf {\bibinfo {volume} {607}},\ \bibinfo {pages}
  {463} (\bibinfo {year} {2022})}\BibitemShut {NoStop}%
\bibitem [{\citenamefont {Anderson}(1958)}]{anderson1958absence}%
  \BibitemOpen
  \bibfield  {author} {\bibinfo {author} {\bibfnamefont {P.~W.}\ \bibnamefont
  {Anderson}},\ }\bibfield  {title} {\bibinfo {title} {Absence of diffusion in
  certain random lattices},\ }\href {https://doi.org/10.1103/PhysRev.109.1492}
  {\bibfield  {journal} {\bibinfo  {journal} {Phys. Rev.}\ }\textbf {\bibinfo
  {volume} {109}},\ \bibinfo {pages} {1492} (\bibinfo {year}
  {1958})}\BibitemShut {NoStop}%
\bibitem [{\citenamefont {Agarwal}\ \emph {et~al.}(2015)\citenamefont
  {Agarwal}, \citenamefont {Gopalakrishnan}, \citenamefont {Knap},
  \citenamefont {M\"uller},\ and\ \citenamefont
  {Demler}}]{agarwal2015anomalous}%
  \BibitemOpen
  \bibfield  {author} {\bibinfo {author} {\bibfnamefont {K.}~\bibnamefont
  {Agarwal}}, \bibinfo {author} {\bibfnamefont {S.}~\bibnamefont
  {Gopalakrishnan}}, \bibinfo {author} {\bibfnamefont {M.}~\bibnamefont
  {Knap}}, \bibinfo {author} {\bibfnamefont {M.}~\bibnamefont {M\"uller}},\
  and\ \bibinfo {author} {\bibfnamefont {E.}~\bibnamefont {Demler}},\
  }\bibfield  {title} {\bibinfo {title} {Anomalous diffusion and griffiths
  effects near the many-body localization transition},\ }\href
  {https://doi.org/10.1103/PhysRevLett.114.160401} {\bibfield  {journal}
  {\bibinfo  {journal} {Phys. Rev. Lett.}\ }\textbf {\bibinfo {volume} {114}},\
  \bibinfo {pages} {160401} (\bibinfo {year} {2015})}\BibitemShut {NoStop}%
\bibitem [{\citenamefont {Martinez}\ and\ \citenamefont
  {Molina}(2006)}]{martinez2006delocalization}%
  \BibitemOpen
  \bibfield  {author} {\bibinfo {author} {\bibfnamefont {D.~F.}\ \bibnamefont
  {Martinez}}\ and\ \bibinfo {author} {\bibfnamefont {R.~A.}\ \bibnamefont
  {Molina}},\ }\bibfield  {title} {\bibinfo {title} {Delocalization induced by
  low-frequency driving in disordered tight-binding lattices},\ }\href
  {https://doi.org/10.1103/PhysRevB.73.073104} {\bibfield  {journal} {\bibinfo
  {journal} {Phys. Rev. B}\ }\textbf {\bibinfo {volume} {73}},\ \bibinfo
  {pages} {073104} (\bibinfo {year} {2006})}\BibitemShut {NoStop}%
\bibitem [{\citenamefont {Abrahams}\ \emph {et~al.}(1979)\citenamefont
  {Abrahams}, \citenamefont {Anderson}, \citenamefont {Licciardello},\ and\
  \citenamefont {Ramakrishnan}}]{abrahams1979scaling}%
  \BibitemOpen
  \bibfield  {author} {\bibinfo {author} {\bibfnamefont {E.}~\bibnamefont
  {Abrahams}}, \bibinfo {author} {\bibfnamefont {P.~W.}\ \bibnamefont
  {Anderson}}, \bibinfo {author} {\bibfnamefont {D.~C.}\ \bibnamefont
  {Licciardello}},\ and\ \bibinfo {author} {\bibfnamefont {T.~V.}\ \bibnamefont
  {Ramakrishnan}},\ }\bibfield  {title} {\bibinfo {title} {Scaling theory of
  localization: Absence of quantum diffusion in two dimensions},\ }\href
  {https://doi.org/10.1103/PhysRevLett.42.673} {\bibfield  {journal} {\bibinfo
  {journal} {Phys. Rev. Lett.}\ }\textbf {\bibinfo {volume} {42}},\ \bibinfo
  {pages} {673} (\bibinfo {year} {1979})}\BibitemShut {NoStop}%
\bibitem [{\citenamefont {Basko}\ \emph {et~al.}(2006)\citenamefont {Basko},
  \citenamefont {Aleiner},\ and\ \citenamefont {Altshuler}}]{basko2006metal}%
  \BibitemOpen
  \bibfield  {author} {\bibinfo {author} {\bibfnamefont {D.~M.}\ \bibnamefont
  {Basko}}, \bibinfo {author} {\bibfnamefont {I.~L.}\ \bibnamefont {Aleiner}},\
  and\ \bibinfo {author} {\bibfnamefont {B.~L.}\ \bibnamefont {Altshuler}},\
  }\bibfield  {title} {\bibinfo {title} {Metal--insulator transition in a
  weakly interacting many-electron system with localized single-particle
  states},\ }\href {https://doi.org/https://doi.org/10.1016/j.aop.2005.11.014}
  {\bibfield  {journal} {\bibinfo  {journal} {Annals of physics}\ }\textbf
  {\bibinfo {volume} {321}},\ \bibinfo {pages} {1126} (\bibinfo {year}
  {2006})}\BibitemShut {NoStop}%
\bibitem [{\citenamefont {{\v{Z}}nidari{\v{c}}}\ \emph
  {et~al.}(2008)\citenamefont {{\v{Z}}nidari{\v{c}}}, \citenamefont {Prosen},\
  and\ \citenamefont {Prelov{\v{s}}ek}}]{vznidarivc2008many}%
  \BibitemOpen
  \bibfield  {author} {\bibinfo {author} {\bibfnamefont {M.}~\bibnamefont
  {{\v{Z}}nidari{\v{c}}}}, \bibinfo {author} {\bibfnamefont {T.}~\bibnamefont
  {Prosen}},\ and\ \bibinfo {author} {\bibfnamefont {P.}~\bibnamefont
  {Prelov{\v{s}}ek}},\ }\bibfield  {title} {\bibinfo {title} {{Many-body
  localization in the Heisenberg XXZ magnet in a random field}},\ }\href
  {https://doi.org/10.1103/PhysRevB.77.064426} {\bibfield  {journal} {\bibinfo
  {journal} {Phys. Rev. B}\ }\textbf {\bibinfo {volume} {77}},\ \bibinfo
  {pages} {064426} (\bibinfo {year} {2008})}\BibitemShut {NoStop}%
\bibitem [{\citenamefont {Luitz}\ \emph {et~al.}(2015)\citenamefont {Luitz},
  \citenamefont {Laflorencie},\ and\ \citenamefont {Alet}}]{luitz2015many}%
  \BibitemOpen
  \bibfield  {author} {\bibinfo {author} {\bibfnamefont {D.~J.}\ \bibnamefont
  {Luitz}}, \bibinfo {author} {\bibfnamefont {N.}~\bibnamefont {Laflorencie}},\
  and\ \bibinfo {author} {\bibfnamefont {F.}~\bibnamefont {Alet}},\ }\bibfield
  {title} {\bibinfo {title} {Many-body localization edge in the random-field
  heisenberg chain},\ }\href {https://doi.org/10.1103/PhysRevB.91.081103}
  {\bibfield  {journal} {\bibinfo  {journal} {Phys. Rev. B}\ }\textbf {\bibinfo
  {volume} {91}},\ \bibinfo {pages} {081103} (\bibinfo {year}
  {2015})}\BibitemShut {NoStop}%
\bibitem [{\citenamefont {Pal}\ and\ \citenamefont {Huse}(2010)}]{pal2010many}%
  \BibitemOpen
  \bibfield  {author} {\bibinfo {author} {\bibfnamefont {A.}~\bibnamefont
  {Pal}}\ and\ \bibinfo {author} {\bibfnamefont {D.~A.}\ \bibnamefont {Huse}},\
  }\bibfield  {title} {\bibinfo {title} {Many-body localization phase
  transition},\ }\href {https://doi.org/10.1103/PhysRevB.82.174411} {\bibfield
  {journal} {\bibinfo  {journal} {Phys. Rev. B}\ }\textbf {\bibinfo {volume}
  {82}},\ \bibinfo {pages} {174411} (\bibinfo {year} {2010})}\BibitemShut
  {NoStop}%
\bibitem [{\citenamefont {{\v{S}}untajs}\ \emph {et~al.}(2020)\citenamefont
  {{\v{S}}untajs}, \citenamefont {Bon{\v{c}}a}, \citenamefont {Prosen},\ and\
  \citenamefont {Vidmar}}]{suntajs2020quantum}%
  \BibitemOpen
  \bibfield  {author} {\bibinfo {author} {\bibfnamefont {J.}~\bibnamefont
  {{\v{S}}untajs}}, \bibinfo {author} {\bibfnamefont {J.}~\bibnamefont
  {Bon{\v{c}}a}}, \bibinfo {author} {\bibfnamefont {T.}~\bibnamefont
  {Prosen}},\ and\ \bibinfo {author} {\bibfnamefont {L.}~\bibnamefont
  {Vidmar}},\ }\bibfield  {title} {\bibinfo {title} {Quantum chaos challenges
  many-body localization},\ }\href
  {https://doi.org/10.1103/PhysRevE.102.062144} {\bibfield  {journal} {\bibinfo
   {journal} {Phys. Rev. E}\ }\textbf {\bibinfo {volume} {102}},\ \bibinfo
  {pages} {062144} (\bibinfo {year} {2020})}\BibitemShut {NoStop}%
\bibitem [{\citenamefont {Abanin}\ \emph {et~al.}(2021)\citenamefont {Abanin},
  \citenamefont {Bardarson}, \citenamefont {{De Tomasi}}, \citenamefont
  {Gopalakrishnan}, \citenamefont {Khemani}, \citenamefont {Parameswaran},
  \citenamefont {Pollmann}, \citenamefont {Potter}, \citenamefont {Serbyn},\
  and\ \citenamefont {Vasseur}}]{abanin2021distinguishing}%
  \BibitemOpen
  \bibfield  {author} {\bibinfo {author} {\bibfnamefont {D.}~\bibnamefont
  {Abanin}}, \bibinfo {author} {\bibfnamefont {J.}~\bibnamefont {Bardarson}},
  \bibinfo {author} {\bibfnamefont {G.}~\bibnamefont {{De Tomasi}}}, \bibinfo
  {author} {\bibfnamefont {S.}~\bibnamefont {Gopalakrishnan}}, \bibinfo
  {author} {\bibfnamefont {V.}~\bibnamefont {Khemani}}, \bibinfo {author}
  {\bibfnamefont {S.}~\bibnamefont {Parameswaran}}, \bibinfo {author}
  {\bibfnamefont {F.}~\bibnamefont {Pollmann}}, \bibinfo {author}
  {\bibfnamefont {A.}~\bibnamefont {Potter}}, \bibinfo {author} {\bibfnamefont
  {M.}~\bibnamefont {Serbyn}},\ and\ \bibinfo {author} {\bibfnamefont
  {R.}~\bibnamefont {Vasseur}},\ }\bibfield  {title} {\bibinfo {title}
  {Distinguishing localization from chaos: Challenges in finite-size systems},\
  }\href {https://doi.org/https://doi.org/10.1016/j.aop.2021.168415} {\bibfield
   {journal} {\bibinfo  {journal} {Annals of Physics}\ }\textbf {\bibinfo
  {volume} {427}},\ \bibinfo {pages} {168415} (\bibinfo {year}
  {2021})}\BibitemShut {NoStop}%
\bibitem [{\citenamefont {Hartmann}\ \emph {et~al.}(2004)\citenamefont
  {Hartmann}, \citenamefont {Keck}, \citenamefont {Korsch},\ and\ \citenamefont
  {Mossmann}}]{hartmann2004dynamics}%
  \BibitemOpen
  \bibfield  {author} {\bibinfo {author} {\bibfnamefont {T.}~\bibnamefont
  {Hartmann}}, \bibinfo {author} {\bibfnamefont {F.}~\bibnamefont {Keck}},
  \bibinfo {author} {\bibfnamefont {H.~J.}\ \bibnamefont {Korsch}},\ and\
  \bibinfo {author} {\bibfnamefont {S.}~\bibnamefont {Mossmann}},\ }\bibfield
  {title} {\bibinfo {title} {{Dynamics of Bloch oscillations}},\ }\href
  {https://doi.org/10.1088/1367-2630/6/1/002} {\bibfield  {journal} {\bibinfo
  {journal} {New Journal of Physics}\ }\textbf {\bibinfo {volume} {6}},\
  \bibinfo {pages} {2} (\bibinfo {year} {2004})}\BibitemShut {NoStop}%
\bibitem [{\citenamefont {Hall}(2015)}]{Hall2015}%
  \BibitemOpen
  \bibfield  {author} {\bibinfo {author} {\bibfnamefont {B.~C.}\ \bibnamefont
  {Hall}},\ }\href {https://doi.org/10.1007/978-3-319-13467-3} {\emph {\bibinfo
  {title} {Lie Groups, Lie Algebras, and Representations}}},\ Vol.\ \bibinfo
  {volume} {222}\ (\bibinfo  {publisher} {Springer International Publishing},\
  \bibinfo {year} {2015})\BibitemShut {NoStop}%
\bibitem [{sup()}]{supplementary}%
  \BibitemOpen
  \bibfield  {title} {\bibinfo {title} {See supplementary for details of
  calculations, additional data, and finite-size effects},\ }\href@noop {} {\
  }\BibitemShut {NoStop}%
\bibitem [{\citenamefont {Nielsen}\ and\ \citenamefont
  {Chuang}(2010)}]{nielsen_chuang_2010}%
  \BibitemOpen
  \bibfield  {author} {\bibinfo {author} {\bibfnamefont {M.~A.}\ \bibnamefont
  {Nielsen}}\ and\ \bibinfo {author} {\bibfnamefont {I.~L.}\ \bibnamefont
  {Chuang}},\ }\href {https://doi.org/10.1017/CBO9780511976667} {\emph
  {\bibinfo {title} {Quantum Computation and Quantum Information: 10th
  Anniversary Edition}}}\ (\bibinfo  {publisher} {Cambridge University Press},\
  \bibinfo {year} {2010})\BibitemShut {NoStop}%
\bibitem [{\citenamefont {Peschel}(2003)}]{peschel2003calculation}%
  \BibitemOpen
  \bibfield  {author} {\bibinfo {author} {\bibfnamefont {I.}~\bibnamefont
  {Peschel}},\ }\bibfield  {title} {\bibinfo {title} {Calculation of reduced
  density matrices from correlation functions},\ }\href
  {https://doi.org/10.1088/0305-4470/36/14/101} {\bibfield  {journal} {\bibinfo
   {journal} {Journal of Physics A: Mathematical and General}\ }\textbf
  {\bibinfo {volume} {36}},\ \bibinfo {pages} {L205} (\bibinfo {year}
  {2003})}\BibitemShut {NoStop}%
\bibitem [{\citenamefont {Roy}\ and\ \citenamefont
  {Sharma}(2018)}]{roy2018entanglement}%
  \BibitemOpen
  \bibfield  {author} {\bibinfo {author} {\bibfnamefont {N.}~\bibnamefont
  {Roy}}\ and\ \bibinfo {author} {\bibfnamefont {A.}~\bibnamefont {Sharma}},\
  }\bibfield  {title} {\bibinfo {title} {Entanglement contour perspective for
  ``strong area-law violation'' in a disordered long-range hopping model},\
  }\href {https://doi.org/10.1103/PhysRevB.97.125116} {\bibfield  {journal}
  {\bibinfo  {journal} {Phys. Rev. B}\ }\textbf {\bibinfo {volume} {97}},\
  \bibinfo {pages} {125116} (\bibinfo {year} {2018})}\BibitemShut {NoStop}%
\bibitem [{Note1()}]{Note1}%
  \BibitemOpen
  \bibinfo {note} {\protect \textit {Since the frequency of the drive
  considered is small, the data corresponding to DL and ADL seem to overlap;
  however, we have checked that for higher frequencies, while a slow relaxation
  is generic, tuning at DL results in a further slowing down of the dynamics
  }\protect \textit {.}}\BibitemShut {Stop}%
\bibitem [{\citenamefont {Hone}\ and\ \citenamefont
  {Holthaus}(1993)}]{hone1993locally}%
  \BibitemOpen
  \bibfield  {author} {\bibinfo {author} {\bibfnamefont {D.~W.}\ \bibnamefont
  {Hone}}\ and\ \bibinfo {author} {\bibfnamefont {M.}~\bibnamefont
  {Holthaus}},\ }\bibfield  {title} {\bibinfo {title} {Locally disordered
  lattices in strong ac electric fields},\ }\href
  {https://doi.org/10.1103/PhysRevB.48.15123} {\bibfield  {journal} {\bibinfo
  {journal} {Phys. Rev. B}\ }\textbf {\bibinfo {volume} {48}},\ \bibinfo
  {pages} {15123} (\bibinfo {year} {1993})}\BibitemShut {NoStop}%
\bibitem [{\citenamefont {Page}(1993)}]{page1993average}%
  \BibitemOpen
  \bibfield  {author} {\bibinfo {author} {\bibfnamefont {D.~N.}\ \bibnamefont
  {Page}},\ }\bibfield  {title} {\bibinfo {title} {Average entropy of a
  subsystem},\ }\href {https://doi.org/10.1103/PhysRevLett.71.1291} {\bibfield
  {journal} {\bibinfo  {journal} {Phys. Rev. Lett.}\ }\textbf {\bibinfo
  {volume} {71}},\ \bibinfo {pages} {1291} (\bibinfo {year}
  {1993})}\BibitemShut {NoStop}%
\bibitem [{\citenamefont {Bar~Lev}\ \emph {et~al.}(2015)\citenamefont
  {Bar~Lev}, \citenamefont {Cohen},\ and\ \citenamefont
  {Reichman}}]{barlev2015absence}%
  \BibitemOpen
  \bibfield  {author} {\bibinfo {author} {\bibfnamefont {Y.}~\bibnamefont
  {Bar~Lev}}, \bibinfo {author} {\bibfnamefont {G.}~\bibnamefont {Cohen}},\
  and\ \bibinfo {author} {\bibfnamefont {D.~R.}\ \bibnamefont {Reichman}},\
  }\bibfield  {title} {\bibinfo {title} {Absence of diffusion in an interacting
  system of spinless fermions on a one-dimensional disordered lattice},\ }\href
  {https://doi.org/10.1103/PhysRevLett.114.100601} {\bibfield  {journal}
  {\bibinfo  {journal} {Phys. Rev. Lett.}\ }\textbf {\bibinfo {volume} {114}},\
  \bibinfo {pages} {100601} (\bibinfo {year} {2015})}\BibitemShut {NoStop}%
\bibitem [{\citenamefont {Luitz}\ and\ \citenamefont
  {Lev}(2017)}]{luitz2017ergodic}%
  \BibitemOpen
  \bibfield  {author} {\bibinfo {author} {\bibfnamefont {D.~J.}\ \bibnamefont
  {Luitz}}\ and\ \bibinfo {author} {\bibfnamefont {Y.~B.}\ \bibnamefont
  {Lev}},\ }\bibfield  {title} {\bibinfo {title} {The ergodic side of the
  many-body localization transition},\ }\href
  {https://doi.org/https://doi.org/10.1002/andp.201600350} {\bibfield
  {journal} {\bibinfo  {journal} {Annalen der Physik}\ }\textbf {\bibinfo
  {volume} {529}},\ \bibinfo {pages} {1600350} (\bibinfo {year}
  {2017})}\BibitemShut {NoStop}%
\bibitem [{\citenamefont {Lezama}\ \emph {et~al.}(2019)\citenamefont {Lezama},
  \citenamefont {Bera},\ and\ \citenamefont {Bardarson}}]{lezama2019apparent}%
  \BibitemOpen
  \bibfield  {author} {\bibinfo {author} {\bibfnamefont {T.~L.~M.}\
  \bibnamefont {Lezama}}, \bibinfo {author} {\bibfnamefont {S.}~\bibnamefont
  {Bera}},\ and\ \bibinfo {author} {\bibfnamefont {J.~H.}\ \bibnamefont
  {Bardarson}},\ }\bibfield  {title} {\bibinfo {title} {Apparent slow dynamics
  in the ergodic phase of a driven many-body localized system without extensive
  conserved quantities},\ }\href {https://doi.org/10.1103/PhysRevB.99.161106}
  {\bibfield  {journal} {\bibinfo  {journal} {Phys. Rev. B}\ }\textbf {\bibinfo
  {volume} {99}},\ \bibinfo {pages} {161106} (\bibinfo {year}
  {2019})}\BibitemShut {NoStop}%
\bibitem [{\citenamefont {Bhakuni}\ \emph {et~al.}(2021)\citenamefont
  {Bhakuni}, \citenamefont {Santos},\ and\ \citenamefont
  {Lev}}]{bhakuni2021suppression}%
  \BibitemOpen
  \bibfield  {author} {\bibinfo {author} {\bibfnamefont {D.~S.}\ \bibnamefont
  {Bhakuni}}, \bibinfo {author} {\bibfnamefont {L.~F.}\ \bibnamefont
  {Santos}},\ and\ \bibinfo {author} {\bibfnamefont {Y.~B.}\ \bibnamefont
  {Lev}},\ }\bibfield  {title} {\bibinfo {title} {Suppression of heating by
  long-range interactions in periodically driven spin chains},\ }\href
  {https://doi.org/10.1103/PhysRevB.104.L140301} {\bibfield  {journal}
  {\bibinfo  {journal} {Phys. Rev. B}\ }\textbf {\bibinfo {volume} {104}},\
  \bibinfo {pages} {L140301} (\bibinfo {year} {2021})}\BibitemShut {NoStop}%
\bibitem [{\citenamefont {Bhakuni}\ \emph {et~al.}(2019)\citenamefont
  {Bhakuni}, \citenamefont {Dattagupta},\ and\ \citenamefont
  {Sharma}}]{bhakuni2019effect}%
  \BibitemOpen
  \bibfield  {author} {\bibinfo {author} {\bibfnamefont {D.~S.}\ \bibnamefont
  {Bhakuni}}, \bibinfo {author} {\bibfnamefont {S.}~\bibnamefont
  {Dattagupta}},\ and\ \bibinfo {author} {\bibfnamefont {A.}~\bibnamefont
  {Sharma}},\ }\bibfield  {title} {\bibinfo {title} {{Effect of noise on Bloch
  oscillations and Wannier-Stark localization}},\ }\href
  {https://doi.org/10.1103/PhysRevB.99.155149} {\bibfield  {journal} {\bibinfo
  {journal} {Phys. Rev. B}\ }\textbf {\bibinfo {volume} {99}},\ \bibinfo
  {pages} {155149} (\bibinfo {year} {2019})}\BibitemShut {NoStop}%
\bibitem [{\citenamefont {Weinberg}\ and\ \citenamefont
  {Bukov}(2017)}]{weinberg2017quspin}%
  \BibitemOpen
  \bibfield  {author} {\bibinfo {author} {\bibfnamefont {P.}~\bibnamefont
  {Weinberg}}\ and\ \bibinfo {author} {\bibfnamefont {M.}~\bibnamefont
  {Bukov}},\ }\bibfield  {title} {\bibinfo {title} {{QuSpin: a Python package
  for dynamics and exact diagonalisation of quantum many body systems part I:
  spin chains}},\ }\href {https://doi.org/10.21468/SciPostPhys.2.1.003}
  {\bibfield  {journal} {\bibinfo  {journal} {SciPost Phys.}\ }\textbf
  {\bibinfo {volume} {2}},\ \bibinfo {pages} {003} (\bibinfo {year}
  {2017})}\BibitemShut {NoStop}%
\bibitem [{\citenamefont {Weinberg}\ and\ \citenamefont
  {Bukov}(2019)}]{weinberg2019quspin}%
  \BibitemOpen
  \bibfield  {author} {\bibinfo {author} {\bibfnamefont {P.}~\bibnamefont
  {Weinberg}}\ and\ \bibinfo {author} {\bibfnamefont {M.}~\bibnamefont
  {Bukov}},\ }\bibfield  {title} {\bibinfo {title} {{QuSpin: a Python package
  for dynamics and exact diagonalisation of quantum many body systems. Part II:
  bosons, fermions and higher spins}},\ }\href
  {https://doi.org/10.21468/SciPostPhys.7.2.020} {\bibfield  {journal}
  {\bibinfo  {journal} {SciPost Phys.}\ }\textbf {\bibinfo {volume} {7}},\
  \bibinfo {pages} {020} (\bibinfo {year} {2019})}\BibitemShut {NoStop}%
\end{thebibliography}%
\ifarXiv
    

\section*{Supplementary material}

# Supplementary material: Dynamical localization and slow dynamics in quasiperiodically-driven quantum systems

Vatsana Tiwari,[1] Devendra Singh Bhakuni,[2] and Auditya Sharma[1, *]

[1] *Department of Physics, Indian Institute of Science Education and Research, Bhopal 462066, India*
[2] *Department of Physics, Ben-Gurion University of the Negev, Beer-Sheva 84105, Israel*

## I. FIBONACCI DRIVING

### A. Non-interacting Limit

For the Fibonacci sequence, we define the unitary operators $U_0 = e^{-iTH_A}$, and $U_1 = e^{-iTH_B}$ corresponding to the electric field $\pm F$ respectively and generate the subsequent evolution according to the Fibonacci sequence [1]:

$$U_n = U_{n-2}U_{n-1}, \quad n \geq 2. \quad (1)$$

Fig. 1(a) shows the Fourier spectrum of the Fibonacci sequence defined as

$$\Gamma_k = \frac{1}{N}\sum_{j=1}^{N} \tau_j \exp\left(\frac{i2\pi k(j-1)}{N}\right), \quad (2)$$

where we calculate the Fourier spectrum using the first $N$ terms of the sequence $\{\tau_j\}$, and $k = 0, 1, 2, ..., N-1$.

For simplicity, let us first study the stroboscopic unitary operator defined at Fibonacci level $m = 2$ with two pulses $A$ and $B$ as

$$U(N=2) = U_B U_A = \exp(-iTH_B)\exp(-iTH_A). \quad (3)$$

Using the BCH formula [2]:

$$e^X e^Y = \exp\left[X + Y + \frac{1}{2}[X,Y] + f([X,Y,[...]]) + ...\right], \quad (4)$$

we get

$$U(N=2) = U_B U_A \equiv \exp(-2iTH_{BA}^{\text{eff}}), \quad (5)$$

where

$$H_{BA}^{\text{eff}} = \Delta_{\text{f}}\left[\hat{K}e^{-i\frac{FT}{2}} + \hat{K}^\dagger e^{i\frac{FT}{2}}\right]. \quad (6)$$

Here, $\Delta_{\text{f}} = \frac{-\Delta}{4}\frac{\sin(FT/2)}{(FT/2)}, T = \frac{2\pi}{\omega}$. Thus, we observe from Eq.(6) that the coefficients of $\hat{K}$ and $\hat{K}^\dagger$ get renormalized. By tuning the amplitude of the drive at $F/\omega = n (n \in \mathbb{Z})$, the effective hopping term vanishes, and

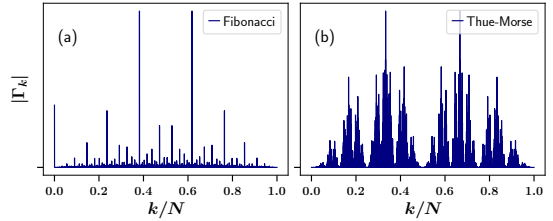

Figure 1. Fourier spectrum of (a) Fibonacci-sequence, and (b) Thue-Morse sequence.

$U(N=2)$ (Eq. (3)) takes the form of an identity operator which implies that any state will return to its initial position after this time period and hence will be dynamically localized. Next, let us consider the case of $m = 4$, where we have 5 pulses: $A, B, A, A, B$. We can define the stroboscopic unitary operator as a product of the operators $U_A$ and $U_B$:

$$U(N=5) = U_B U_A U_A U_B U_A. \quad (7)$$

It is useful to pair up consecutive unitary operators associated with different types of pulses ($A$ and $B$); we thus define the Floquet unitary operators $U_C$ and $U_{C'}$:

$$U_C = U_B U_A \equiv \exp(-2iTH_{BA}^{\text{eff}}), \quad (8)$$

$$H_{BA}^{\text{eff}} = \Delta_{\text{f}}\left[\hat{K}e^{-i\frac{FT}{2}} + \hat{K}^\dagger e^{i\frac{FT}{2}}\right], \quad (9)$$

$$U_{C'} = U_A U_B \equiv \exp(-2iTH_{AB}^{\text{eff}}), \quad (10)$$

$$H_{AB}^{\text{eff}} = \Delta_{\text{f}}\left[\hat{K}e^{i\frac{FT}{2}} + \hat{K}^\dagger e^{-i\frac{FT}{2}}\right]. \quad (11)$$

Using Eq. (8), and (10), we can rewrite Eq. (7) as:

$$\begin{aligned}U(N=5) &= U_C U_{C'} U_A, \\ &\equiv \exp(-2iTH_{BA}^{\text{eff}})\exp(-2iTH_{AB}^{\text{eff}})U_A.\end{aligned} \quad (12)$$

It is clear from Eq. (9) and Eq. (11) that at $F = n\omega$, both $U_C$ and $U_{C'}$ become the identity operator and the effective dynamics at Fibonacci level $m = 4$ is governed by $U_A$:

$$U(N=5) \equiv U_A = \exp(-iTH_A). \quad (13)$$

A careful analysis of the Fibonacci sequence at higher levels suggests that the product of two different consecutive unitary operators ($U_A$ and $U_B$) yields an identity

* auditya@iiserb.ac.in

operator at the dynamical localization point $F = n\omega$, and we are left with a string of $U_A$s with length $n_q$, where $n_q$ are Fibonacci numbers. Hence, we obtain an effective Floquet operator given by $(U_A)^{n_q}$. We can write this out explicitly as:

$$U(m=2) \equiv (U_A)^{n_1} = \exp(-in_1 TH_A) \quad ,$$
$$U(m=3) \equiv (U_A)^{n_2} = \exp(-in_2 TH_A) \quad ,$$
$$U(m=4) \equiv (U_A)^{n_3} = \exp(-in_3 TH_A) \quad ,$$
$$\ldots\ldots\ldots\ldots\ldots\ldots\ldots\ldots\ldots\ldots\ldots\ldots\ldots \quad ,$$
$$U(m) \equiv (U_A)^{n_{m-1}} = \exp(-in_{m-1} TH_A) \quad , \quad (14)$$

where $n_q = 0, 1, 1, 2, \ldots$ are Fibonaaci numbers. Hence, we can conclude that the dynamics is governed by the static Hamiltonian $H_A$ at the dynamical localization point $\frac{F}{\omega} = n$. This comes from the fact that we are starting our protocol from $U_0$ governed by $H_A$. If instead we start from $H_B$, then the effective unitary operator would be controlled by $U_B$ corresponding to the static Hamiltonian $H_B$. Since the dynamics is entirely controlled by a single static Hamiltonian, the dynamical localization we see here is effectively just Bloch oscillations corresponding to the static Hamiltonian.

We verify these analytical results with the help of the dynamics of return probability $P(t)$ and entanglement entropy $S(t)$ in Fig. (2)(a,c). Further to study the behavior of the disordered non-interacting system under Fibonacci driving, we again study the return probability and entanglement entropy. Fig. (2)(b,d) show that dynamical localization is destroyed and entagement entropy grows with a sublinear power law, $S(t) \propto t^\gamma, \gamma < 1$. This is qualitatively similar to the Thue-Morse driven case.

In Fig. (3), we show the dynamics of return probability $P(t)$ and entanglement entropy $S(t)$ for a Fibonacci driven disordered system for various values of disorder strength $W$ and frequency $\omega$. For a particular disorder strength, the dynamics slows down as the frequency increases. On the other hand, comparing $P(t)$ in Fig. 3(a) and Fig. (3)(b), we find that the return probability decays faster for disorder strength $W = 3$ (Fig 3(b)) at certain values of the frequency. Such a feature is consistent with the argument presented in the main paper that driving frequencies larger than the local bandwidth $\Delta_\xi$ corresponding to a localization length $\xi$, do not affect localization and driving frequencies smaller than that will lead to the development of an overlap between the localized states. A competition between these high and low frequency components leads to a slow relaxation. For frequencies $\omega = 1, 3$, we observe that the data presented for disorder strengths $W = 1$ and $W = 3$ are very similar. On the other hand, for frequenices $\omega = 5$ and higher, the timescales involved for $W = 1$ are much higher than those for $W = 3$. Thus, we see that when the disorder strength is higher, a higher frequency is needed to observe slow relaxation.

Figure 2. (a,b): Dynamics of return probability $P(t)$ at dynamical localization point ($F = n\omega$) and away from dynamical localization($F \neq n\omega$) for Fibonacci driven clean ($W = 0.0$) and disordered non-interacting system. (c,d): Entanglement entropy $S(t)$ for Fibonacci driven clean ($W = 0.0$) and disordered ($W = 5.0$) non-interacting system at dynamical localization point ($F = n\omega$) and away from dynamical localization($F \neq n\omega$) point. The other parameters are $\Delta = 4.0, \omega = 1.0, L = 200$. The time discretization for our numerics is $\Delta t = 0.001$. For the disordered case, we have averaged over 100 realizations of disorder. In (d), the (linear) black dashed line indicates the ballistic growth of entanglement entropy ($S(t) \propto t$).

Figure 3. (a,b). Dynamics of return probability $P(t)$ and (c,d) Half chain entanglement entropy $S(t)$ for Fibonacci driven disordered ($W = 1.0, 3.0$) non-interacting system for various values of driving-frequency $\omega$ and amplitude $F$. The other parameters are $\Delta = 4.0, F = 2.0\omega, U = 0.0$.

## II. THUE-MORSE DRIVING

### A. Non-interacting Limit

The Thue-Morse sequence (TMS) can be generated using the recurrence relation [3]:

$$U_{n+1} = \tilde{U}_n U_n, \quad \tilde{U}_{n+1} = U_n \tilde{U}_n, \quad (15)$$

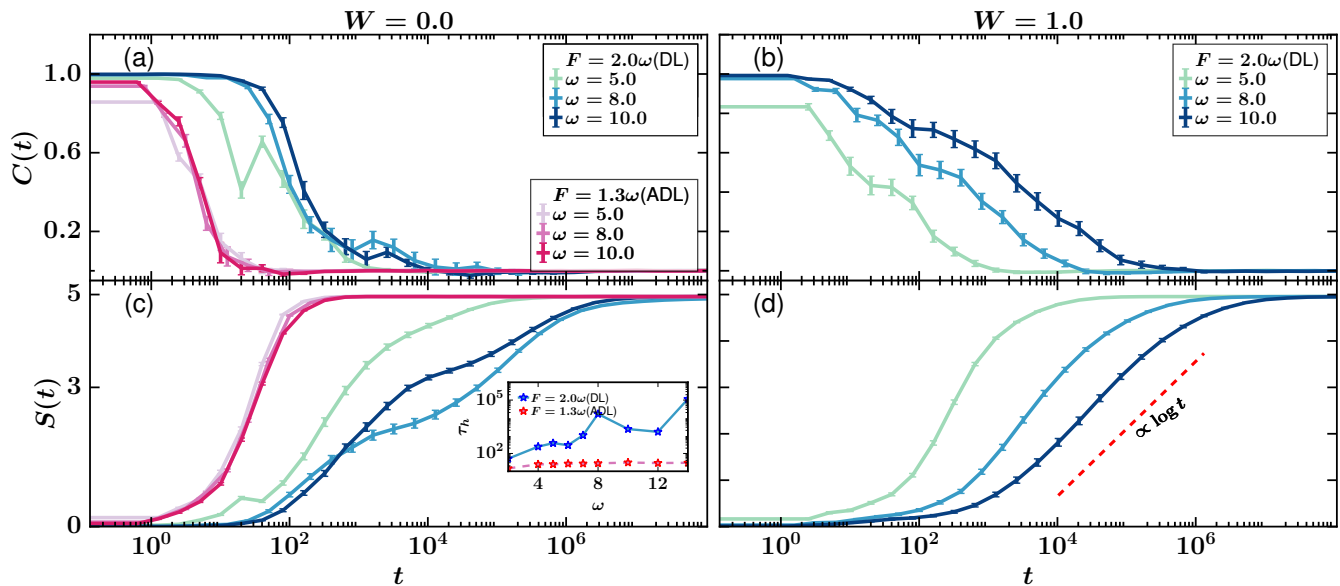

Figure 4. (a,c): Auto-correlation $C(t)$ and half chain entanglement entropy $S(t)$ for Thue-Morse driven clean ($W = 0.0$) interacting system. In (a) and (c), the plots in pink shades correspond to the parameters being tuned away from dynamical localization (ADL) ($F = 1.3\omega$), and blue shade plots correspond to the parameters being tuned at dynamical localization (DL) ($F = 2.0\omega$). Inset(c): Heating time $\tau_h$ vs frequency $\omega$. (b, e): $C(t)$ and $S(t)$ for Thue-Morse driven disordered ($W = 1.0$) interacting system. We have averaged over 100 disorder realizations. In (d), the red dashed line is drawn to show logarithmic growth of entanglement entropy ($S(t) \propto \log t$) for $\omega = 10$. The other system parameters are $\Delta = 4.0, U = 1.0, F = 2.0\omega$, and system size $L = 16$.

where we start with the unitary operators $U_1 = U_B U_A$, $\tilde{U}_1 = U_A U_B$ with $U_A = e^{-iTH_A}$ and $U_B = e^{-iTH_B}$. Fig. (1)(b) shows the Fourier spectrum of the Thue-Morse sequence.

To study the dynamics of Thue-Morse driven systems, we evaluate the Floquet unitary operator stroboscopically. Let us first compute the time evolution operator at Thue-Morse level $m = 2$, where the drive comes with $4(= 2^2)$ pulses: $A, B, B, A$. Hence, we can write the time evolution operator stroboscopically as:

$$U(N = 4) = U_A U_B U_B U_A. \quad (16)$$

From the recurrence relation (15), we see that the Thue-Morse sequence always yields pairs of $U_A$ and $U_B$, with the total number of these operators being an exact power of 2. Hence, we can evaluate Eq.(16) as

$$U(N = 4) = U_{C'} U_C, \quad (17)$$

where the operators $U_C$ and $U_{C'}$ are as defined earlier:

$$U_C = U_B U_A \equiv \exp(-2iTH_{BA}^{\text{eff}}), \quad (18)$$

$$H_{BA}^{\text{eff}} = \Delta_f \left[ \hat{K} e^{-i\frac{FT}{2}} + \hat{K}^\dagger e^{i\frac{FT}{2}} \right], \quad (19)$$

$$U_{C'} = U_A U_B \equiv \exp(-2iTH_{AB}^{\text{eff}}), \quad (20)$$

$$H_{AB}^{\text{eff}} = \Delta_f \left[ \hat{K} e^{i\frac{FT}{2}} + \hat{K}^\dagger e^{-i\frac{FT}{2}} \right]. \quad (21)$$

Using the BCH formula(4), we can rewrite Eq. (17) in terms of an effective Hamiltonian as

$$U(N = 2^2) \equiv \exp\left(-2^2 iTH_{\text{eff}}\right), \quad (22)$$

where

$$H_{\text{eff}} = \Delta_{\text{eff}} \left( \hat{K} + \hat{K}^\dagger \right), \quad \Delta_{\text{eff}} = \frac{-\Delta}{4} \frac{\sin FT}{FT}. \quad (23)$$

Here, $\Delta_{\text{eff}}$ is the effective hopping strength which vanishes when the drive strength $F$ is tuned to be a half integer multiple of drive frequency $\omega$ i.e. $F/\omega = n/2, n \in \mathbb{Z}$. At these specific values, Thue-Morse driving causes the dynamical localization of the non-interacting system.

Following the recurrence relation defined in Eq. (15), we can construct the time evolution operator $U(N = 2^m)$ for higher Thue-Morse level $m$ having $2^m$ pulses, which is the product of a string of Floquet unitary operators $U_A$ and $U_B$. We observe that the full product of unitary operators at each Thue-Morse level $m > 2$ may be constructed in terms of groups of four operators of two types: '$U_A U_B U_B U_A$' and '$U_B U_A U_A U_B$'. Since $U_{C'} U_C \equiv U_C U_{C'}$ (from Eq. (18)-(20)), following the steps from Eq. (16)-Eq. (22), we can again get a generalized form of the unitary operator $U(N = 2^m)$:

$$U(N = 2^m) \equiv \left(\exp(-2^2 iTH_{\text{eff}})\right)^{2^{m-2}}$$
$$U(N = 2^m) \equiv \exp\left(-2^m iTH_{\text{eff}}\right). \quad (24)$$

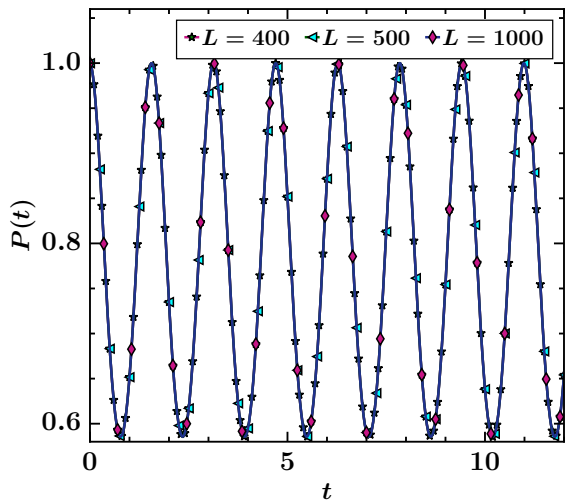

Figure 5. Dynamics of return probability $P(t)$ at dynamical localization point ($F = 2\omega$) for Thue-Morse driven clean($W = 0$) non-interacting system for system sizes $L = 400, 500, 1000$. The other parameters are $\Delta = 4.0, \omega = 2.0, F = 2.0, U = 0.0$.

Hence, we observe that the dynamics of the Thue-Morse driven system is governed by $H_{\text{eff}}$ which is simply a nearest-neighbor tight-binding Hamiltonian with renormalized hopping strength $\Delta_{\text{eff}}$ (Eq. (23)). In the main Letter, we show numerical evidence for this analytical result in the form of return probability $P(t)$ and entanglement entropy $S(t)$. Next, we include disorder in addition to the Thue-Morse drive. The disorder destroys the dynamical localization and we observe a sublinear growth of entanglement entropy, $S(t) \propto t^\gamma, \gamma < 1$ as shown in the main Letter (Fig.(1) of the main Letter).

Finally, we comment on the possibility of dynamical localization for any sequence of random unitary operators [3]. While the system is delocalized when the electric field is randomly flipped at arbitrary times (white-noise) [4], dynamical localization can be engineered for any random sequence of unitary operators applied for an equal amount of time and with proper tuning of the parameters. This can be accomplished by a method similar to the above where a proper choice of the parameters can yield an effective unitary operator that is equal to the identity operator.

### B. Interacting Limit

We study the behavior of the auto-correlation function $C(t) = 4\langle \hat{S}^z_{L/2}(t)\hat{S}^z_{L/2}(0)\rangle$ and growth of entanglement entropy $S(t)$ for the Thue-Morse driven clean (Fig. 4(a,c)) and disordered interacting (Fig. 4(b,d)) system. For the clean system, the dynamics for $F/\omega = 1.3$ (away from dynamical localization) quickly approaches the Page value (pink shade curves), whereas for $F/\omega = $ 2.0, the system relaxes to the infinite temperature state very slowly (blue shade curves). To distinguish the dynamics of the system at the dynamical localization (DL) point and away from the dynamical localization (ADL) point, we plot the heating time as a function of frequency ($\tau_h$ vs $\omega$) in the inset of Fig.(4(c)) for $F = 2.0\omega$ (DL), and $F = 1.3\omega$ (ADL). It clearly shows that the significant suppression of hopping strength at DL leads to a slow heating of the system. In contrast, the system quickly reaches a featureless infinite temperature state on tuning the parameters away from DL.

In Fig. 4(b,d), we focus on the ergodic side of the disordered system by fixing the disorder strength to be $W = 1.0$, and drive the system at the dynamical localization point $F/\omega = 2.0$. $C(t)$ and $S(t)$ exhibit the same behavior as shown in the Fibonacci driven case in the main Letter. The entanglement entropy attains the Page value, $S_{\text{Page}} = \frac{1}{2}(L \ln 2 - 1)$ ( corresponding to the infinite temperature state) [5] logarithmically slowly ($S(t) \propto \log t$) (Fig. 4(d)).

### III. FINITE-SIZE EFFECTS

To show the system size independence of our chief findings for driven non-interacting and interacting systems, we show the data for different system sizes in Fig. 5, 6 and Fig. 7. In the non-interacting clean limit, we study return probability $P(t)$ at the dynamical localization point for different system sizes $L = 400, 500, 1000$ in Fig. 5, and find that the dynamics of return probability remains consistent for all the considered system sizes. For the non-interacting disordered system, we study the return probability $P(t)$, and half chain entanglement entropy $S(t)$ for various system sizes $L = 100, 200, 400, 600$. In Fig. 6(a) and Fig. 6(b), the return probability $P(t)$ and half chain entanglement entropy $S(t)$ exhibit slow relaxation for all the studied system sizes. For the interacting model, we study auto-correlation function $C(t)$, and half chain entanglement entropy $S(t)$ for system sizes $L = 10, 12, 14$ and show the data for clean and disordered cases in Fig.( 7(a,c)) and Fig.( 7(b,d)). In both the cases, we observe that the dynamics of entanglement entropy for small system sizes exhibits slow relaxation to the Page value $S_{\text{Page}} = \frac{1}{2}(L \ln 2 - 1)$ corresponding to an infinite-temperature state, and hence is in complete agreement with the findings for system size $L = 16$ (as shown in the main Letter). The dynamics of autocorrelation function $C(t)$ also exhibits slow relaxation for all the studied system sizes. Thus, we can conclude that the quasiperiodic drive induces slow relaxation in the non-interacting and interacting disordered system. While $S(t)$ follows sublinear growth ($S(t) \propto t^\gamma, \gamma < 1$) in the non-interacting case, we observe logarithmic growth of entanglement entropy, $S(t) \propto \log t$ in the interacting case. We have also checked the results for the Thue-Morse driven system and confirm that this too yields similar behavior.

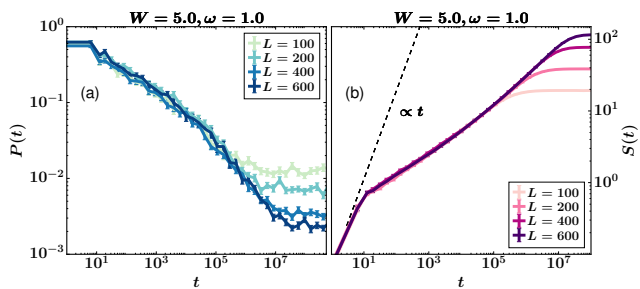

Figure 6. (a) Dynamics of return probability $P(t)$ and (b) Half chain entanglement entropy $S(t)$ for Fibonacci driven disordered ($W = 5.0$) non-interacting system for system sizes $L = 100, 200, 400, 600$. The black dashed line indicates the linear growth of entanglement entropy ($S(t) \propto t$). The other parameters are $\Delta = 4.0, \omega = 1.0, F = 2.0, U = 0.0$.

## IV. PERIODICALLY DRIVEN MODEL

### A. Non-interacting Limit

To understand the dynamics of quasiperiodically driven systems, we start with the analysis of a periodically driven system defined by the Hamiltonian

$$H(t) = \begin{cases} H_A = D + D_0 + E, & 0 \leq t \leq T/2 \\ H_B = D + D_0 - E, & T/2 < t \leq T, \end{cases} \quad (25)$$

where,

$$D = -\frac{\Delta}{4} \sum_{j=0}^{L-2} \left( c_j^\dagger c_{j+1} + h.c. \right) = \frac{-\Delta}{4}(\hat{K} + \hat{K}^\dagger),$$

$$D_0 = \sum_{j=0}^{L-1} h_j \left( n_j - \frac{1}{2} \right),$$

$$E = F \sum_{j=0}^{L-1} j \left( n_j - \frac{1}{2} \right) = F\hat{N}. \quad (26)$$

Now, we define the time evolution operator $U = U_A U_B$ and extract the effective Hamiltonian $H_{\text{eff}}$:

$$U = U_A U_B = \exp(-i\frac{T}{2}H_A)\exp(-i\frac{T}{2}H_B) \equiv \exp(-iTH_{\text{eff}}) \quad (27)$$

where $H_{\text{eff}}$ is defined as

$$H_{\text{eff}} = \frac{-1}{iT} \log\left[\exp(-i\frac{T}{2}H_A)\exp(-i\frac{T}{2}H_B)\right],$$
$$= H_\Delta + D_0 + H_{LRH}.$$

Using the BCH expansion [2] we can work out

$$H_\Delta = \Delta_{\text{eff}}\{\hat{K}e^{-iFT/4} + \hat{K}^\dagger e^{iFT/4}\}, \quad (28)$$

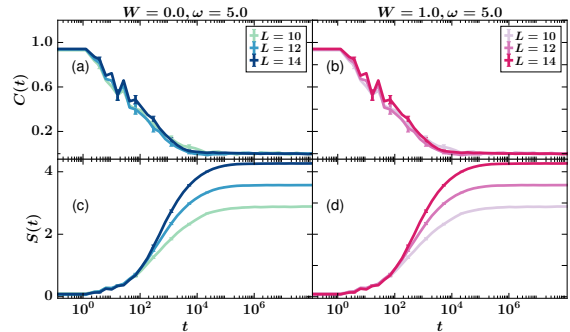

Figure 7. Comparison of dynamics of Fibonacci driven clean ($W = 0.0$) and disordered ($W = 1.0$) interacting system. (a,b): Auto-correlation function $C(t)$ at driving frequency $\omega = 5.0$ for different system sizes $L = 10, 12, 14$. (c,d): Half chain entanglement entropy $S(t)$ at driving frequency $\omega = 5.0$ for different system sizes $L = 10, 12, 14$. The other parameters are $\Delta = 4.0, F = 2\omega, U = 1.0$.

with $\Delta_{\text{eff}} = \frac{-\Delta/4}{F\pi/2\omega} \sin(F\pi/2\omega)$ as before. $H_{LRH}$ contains the longer-range hopping terms and for which we obtain an approximate expression (by considering upto 5th order terms in the BCH expansion):

$$H_{LRH} = \frac{-\Delta}{4} FT \sum_j \Big[ d_j(a + ib)\left(c_{j+2}^\dagger c_j + h.c.\right) + \{(q_1 b_j^2 + q_2 b_j) + i(p_1 b_j^2 + p_2 b_j)\} \left(c_{j+1}^\dagger c_j + h.c.\right) + b_j\left(c_{j+1}^\dagger c_{j+1} - c_j^\dagger c_j\right)\Big], \quad (29)$$

where $d_j = b_{j+1} - b_j, b_j = h_{j+1} - h_j$, and $a = \frac{-13T^3(\Delta F)}{11520}$, $b = \frac{-\Delta T^2}{384}$, $q_1 = \frac{-7FT^3}{1440}$, $p_1 = \frac{-T^2}{48}$, $q_2 = \frac{-F^2 T^3}{480}$, $p_2 = \frac{-T^2 F}{96}(F - 1)$.

#### 1. High-frequency Driving

If driving frequency $\omega$ is larger than the local bandwidth $\Delta_\xi$ of a system of size of the order of the localization length $\xi$ (it implies that $\omega >> \Delta, W$), we can ignore the terms with higher power of $T(= 2\pi/\omega)$. Hence, we recover an effective Hamiltonian $H_{\text{eff}}$:

$$H_{\text{eff}} = H_\Delta + D_0. \quad (30)$$

At dynamical localization point, $F = 2n\omega$, $\Delta_{\text{eff}} = \frac{-\Delta/4}{F\pi/2\omega}\sin(F\pi/2\omega) \to 0$, and we get hopping suppression from $\Delta$ to $\Delta_{\text{eff}}$. It leads to an increased normalized disorder strength from $W/\Delta$ to $W/\Delta_{\text{eff}}$. Thus, the periodically driven model shows the persistence of *Anderson localization* when the drive-parameters are tuned at the dynamical localization (DL) point at high frequency [6, 7].

*Away from dynamical localization (ADL)*, if we again analyse $\Delta_{\text{eff}}$ at high frequency:

$$\Delta_{\text{eff}} = \frac{-\Delta/4}{F\pi/2\omega}\sin(F\pi/2\omega) \equiv \frac{\Delta}{4}; \quad (\Delta/\omega << 1),$$
$$H_{\text{eff}} = \frac{-\Delta}{4}\left(\hat{K}e^{-iFT/4} + \hat{K}^{\dagger}e^{iFT/4}\right) + D_0. \quad (31)$$

Hence, we recover an Anderson-like model Hamiltonian at high frequency driving even at points away from DL. Therefore, we observe that high-frequency periodic driving preserves Anderson localization even away from DL.

### 2. Low Frequency Driving

However, at low-frequency when driving frequency $\omega$ is smaller than the local bandwidth $\Delta_\xi$ (i.e. $\omega << \Delta, W$), $H_{LRH}$ (Eq.(29)) which contains the higher order hopping terms of the effective Hamiltonian plays an important role. This causes delocalization at DL as well as away from DL.

### B. Interacting Limit

The Hamiltonian here is:

$$H(t) = \begin{cases} H_A = D + D_0 + V + E, & 0 \leq t \leq T/2 \\ H_B = D + D_0 + V - E, & T/2 < t \leq T, \end{cases}$$
$$V = \sum_{j=0}^{L-2} U(n_j - \frac{1}{2})(n_{j+1} - \frac{1}{2}).$$

Following the analysis of the non-interacting disordered Floquet Hamiltonian, we can again write an effective Hamiltonian:

$$H_{\text{eff}} = H_\Delta + D_0 + V + H_{LR}, \quad (32)$$

where

$$H_{LR} = H_{LRH} + H_{int}(V, D, D_0, E), \quad (33)$$

and

$$H_{int}(V, D, D_0, E) = -\frac{(-iT)^2}{12}\{[V,[H_A,[E,D]]] \\ +[H_B, A] + [D + D_0 + V, A]\} + ..., \quad (34)$$

where $A = [V, [E, D]]$. For the case of the interacting Hamiltonian, we obtain $H_{LR}$ by replacing $D_0$ with $D_0+V$ in the computation of the non-interacting limit. Again it is worth noting that $H_{LR}$ comes with higher powers of $T$.

#### 1. High-frequency Driving

Similar to the non-interacting limit, we can ignore $H_{LR}$ at high-frequency $\omega(>> \Delta)$, and get an effective Hamiltonian as the sum of $H_\Delta$, $D_0$, and $V$:

$$H_{\text{eff}} = H_\Delta + D_0 + V, \quad (35)$$

which is in the form of a *standard MBL model, with hopping strength* $\Delta_{\text{eff}}$.

At DL, $\Delta_{\text{eff}} = \frac{-\Delta/4}{F\pi/2\omega}\sin(F\pi/2\omega) \to 0$, and normalized disorder increases from $\frac{W}{\Delta}$ to $\frac{W}{\Delta_{\text{eff}}}$, hence leads to *drive-induced MBL* [8]. At this frequency, the system is endowed with a set of local integrals of motion which freeze transport and localize excitations. It is worth noting that $\omega_c$, the threshold value above which localization occurs, varies with $W$.

*ADL*, from a series expansion of $\sin c$ function:

$$\Delta_{\text{eff}} = \frac{-\Delta/4}{F\pi/2\omega}\sin(F\pi/2\omega) \equiv \frac{\Delta}{4}. \quad (36)$$

Hence,

$$H_{\text{eff}} = \frac{-\Delta}{4}\{\hat{K}e^{-iFT/4} + \hat{K}^{\dagger}e^{iFT/4}\} + D_0 + V. \quad (37)$$

Eq.(37) is again an *MBL-like* Hamiltonian.

[1] S. Maity, U. Bhattacharya, A. Dutta, and D. Sen, Fibonacci steady states in a driven integrable quantum system, Phys. Rev. B **99**, 020306 (2019).
[2] B. C. Hall, *Lie Groups, Lie Algebras, and Representations*, Vol. 222 (Springer International Publishing, 2015).
[3] H. Zhao, F. Mintert, R. Moessner, and J. Knolle, Random multipolar driving: Tunably slow heating through spectral engineering, Phys. Rev. Lett. **126**, 040601 (2021).
[4] V. Tiwari, D. S. Bhakuni, and A. Sharma, Noise-induced dynamical localization and delocalization, Phys. Rev. B **105**, 165114 (2022).
[5] D. N. Page, Average entropy of a subsystem, Phys. Rev. Lett. **71**, 1291 (1993).
[6] D. W. Hone and M. Holthaus, Locally disordered lattices in strong ac electric fields, Phys. Rev. B **48**, 15123 (1993).
[7] D. F. Martinez and R. A. Molina, Delocalization induced by low-frequency driving in disordered tight-binding lattices, Phys. Rev. B **73**, 073104 (2006).
[8] E. Bairey, G. Refael, and N. H. Lindner, Driving induced many-body localization, Phys. Rev. B **96**, 020201 (2017).


\fi

\end{document}